\documentclass[%
 aip,
 amsmath,amssymb,
 reprint,%
]{revtex4-1}
 
\usepackage{graphicx}
\usepackage{dcolumn}
\usepackage{bm}

\usepackage[utf8]{inputenc}
\usepackage[T1]{fontenc}
\usepackage{mathptmx}
\usepackage{etoolbox}

\usepackage{makecell, booktabs, multirow, siunitx}
\usepackage[normalem]{ulem}
\usepackage{xcolor,soul,framed} 

\makeatletter
\def\@email#1#2{%
 \endgroup
 \patchcmd{\titleblock@produce}
  {\frontmatter@RRAPformat}
  {\frontmatter@RRAPformat{\produce@RRAP{*#1\href{mailto:#2}{#2}}}\frontmatter@RRAPformat}
  {}{}
}%
\makeatother

\begin{document}

\preprint{AIP/123-QED}

\title[ ]{Scaling Up Silicon Photonic-based Accelerators:\\ Challenges and Opportunities}
\author{M. A. Al-Qadasi}
\affiliation{ 
The  University  of  British  Columbia,Vancouver, BC V6T 1Z4, Canada. 
}%
\author{L. Chrostowski}%
\affiliation{ 
The  University  of  British  Columbia,Vancouver, BC V6T 1Z4, Canada. 
}%
\affiliation{ 
Stewart Blusson Quantum Matter Institute, University of British Columbia, Vancouver, BC V6T 1Z4, Canada
}
\author{B. J. Shastri}%
\affiliation{Queen's University, Kingston, ON K7L 3N6, Canada.}
\affiliation{Vector Institute, Toronto, ON M5G 1M1, Canada.}
\author{S. Shekhar}%
 \affiliation{ 
The  University  of  British  Columbia,Vancouver, BC V6T 1Z4, Canada. 
}%

\date{09 September 2021}

\begin{abstract}
Digital accelerators in the latest generation of CMOS processes support multiply and accumulate (MAC) operations at energy efficiencies spanning 10-to-100~fJ/Op. But the operating speed for such MAC operations are often limited to a few hundreds of MHz.
Optical or optoelectronic MAC operations on today's SOI-based silicon photonic integrated circuit platforms can be realized at a speed of tens of GHz, leading to much lower latency and higher throughput. In this paper, we study the energy efficiency of integrated silicon photonic MAC circuits based on Mach-Zehnder modulators and microring resonators. We describe the bounds on energy efficiency and scaling limits for $N\times N$ optical networks with today's technology, based on the optical and electrical link budget. We also describe research directions that can overcome the current limitations. 
\end{abstract}

\maketitle

\section{\label{sec:level1}Introduction:\protect\\ }

Vector matrix multiplication operations represents the core of artificial neural networks (ANNs) and other computing applications of hardware accelerators. ANNs are realized in digital complementary metal–oxide-semiconductor (CMOS) circuits with multiple processing elements implementing multiply and accumulate (MAC) operations which  calculates the product of two numbers and adds the result to an accumulator \cite{sze_efficient_processing_of_DNN}. The processing elements can be arranged in a systolic architecture, where data is passed through connected processing elements in a rhythmic sequence, to perform MACs either spatially or temporally over several clock cycles \cite{Kung_systolic_architectures}. 

Integrated silicon photonics (SiP) circuits have been popularly employed in high-speed links to move data at a rate of tens of Gb/s, where optical modulation is more efficient than electronic switching for transmitting data over significant distances \cite{AR_JSSC_A_Dual_Polarization_Silicon_Photonic_Coherent_Transmitter_Supporting_552_Gbs_wavelength}. Optical modulation can be realized using Mach Zehnder modulators (MZMs) or microring modulators (MRMs) \cite{AR_Silicon_Photonics_MRR_Datacenters}. MZMs are broadband and easily support complex modulation schemes \cite{AR_JSSC_A_Dual_Polarization_Silicon_Photonic_Coherent_Transmitter_Supporting_552_Gbs_wavelength}. MRMs have significantly smaller footprint and driver power consumption \cite{Sajjad_moazeni_Vladimir_A_40Gbps_PAM4_Tx_based_on_a_ring_resonator_optical_DAC_in45nm_SOI_CMOS}. As a technology, the current generation of SiP has now matured with high volume shipments for datacenter transceivers from companies such as Intel and Cisco. SiP circuits comprising of Mach Zehnder interferometers (MZIs) or microring resonators (MRRs) have been used also for other applications such as high-speed optical switches and filters \cite{AlTaha_Monitoring_and_automatic_tuning_and_stabilization_of_2x2_mzi_optical_switch, Hasitha_Jayatilleka_Photoconductive_heaters_enable_control_of_large_scale_silicon_photonic_ring_resonator_circuits, Po_Dong_Commercializing_Silicon_Microring_Resonators_Technical_Challenges_and_Potential_Solutions,Kazuhiro_Ikeda_Large_scale_Silicon_photonics_switch, Bhaskar_Englund_Memory_enhanced_quantum_communication}.

SiP is also being used for computing applications \cite{Tait_2019, shen_2017, hughes_2018, Viraj_Bangari_Digiral_Electronics_and_Analog_Photonics_DEAP_CNN,Shekhar2021_ISSCMagazine,Carl_Ramey_Lightmatter_Hotchips_Silicon_Photonics_for_Artificial_Intelligence_Acceleration}, where devices such as MZIs, MZMs, MRRs, and MRMs are used for computation in optical analog domain. These encompass inference and training accelerators used for machine learning and neuromorphic computing applications where convolution takes $80\%$ of the total processing time \cite{Li_GPU_2016, Chaoran_2022, filipovich2021monolithic}. Other integrated optical configurations implemented using field-programmable photonic arrays were shown to carry out linear transformations for signal  processing and control \cite{Perez_2018, Zhang_yao_2020}. Linear transformation circuits are also employed in Ising machines\cite{Nikolas2019_IsingMachine} and photonic quantum computing processors \cite{Qiang_2018}.


In this Perspective, we describe the advantages and challenges of implementing MACs using SiP, and comment on how to address them. The paper is organized as follows: Section II and III describe the link budget, energy efficiency and scaling opportunities for SiP MACs implemented with MZM and MRM, respectively. Section IV explores the possible approaches to further improve SiP MAC systems. It introduces the ongoing research in the field of SiP that when fully realized, will lead to significant changes in the field of optical computing and communication. Section V concludes the paper.

\section{MZM Based Si-photonic Implementation}
\label{section_MZM_based_SiP_Implementation}
\subsection{System Architecture}

Fig.~\ref{schematic_8N} illustrates an MZM-based SiP implementation of an optical accelerator. Using off-chip lasers, light is guided by a polarization-maintaining (PM) single mode fiber (SMF), gets coupled to the SiP chip via an edge coupler and then split to $N$ parts. These parts are modulated by an array of $N$ MZI modulators and fed into an $N\times N$ weight transformation (multiplication) matrix, $W_{N\times N}$. The optical intensities at the MAC outputs, $Y_{N\times 1}$, are thus described by the multiplication product of the input vector, $V_{N\times 1}$ and the weight matrix, $W_{N\times N}$, as:
\begin{equation}
  Y_{N\times 1}=W_{N\times N}V_{N\times 1}
\end{equation}

In other words, the weight matrix performs linear transformation for the input vector, $V_{N\times 1}$, and delivers $N$ outputs, $Y_{N\times 1}$, that are routed to an array of photodetectors (PDs). These  PDs  are  then connected to the electrical components, including TIAs, main amplifiers, and sense amplifiers based comparators, which are either built in a separate CMOS/BiCMOS chip, or monolithically integrated with the SiP devices in the same process.  

Singular value decomposition (SVD) is an effective approach to represent a given matrix as a factorization of multiple matrices \cite{svd_1995,SVD_1970}. SVD decomposes a real matrix into a product of unitary matrices and a diagonal matrix. This is useful in the experimental realization of an $N\times N$ matrix topology in which the sequential product of rotation matrices represents the sequential arrangement of linear transformation units in the overall matrix grid \cite{reck}. A $2\times 2$ linear transformation unit in the whole grid arrangement is practically implemented using a tunable beam splitter (TBS) as seen in Fig.~\ref{schematic_8N} \cite{clements, AlTaha_Monitoring_and_automatic_tuning_and_stabilization_of_2x2_mzi_optical_switch}. A TBS comprises of an MZI with a phase shifter ($\theta$) in at least one of the arms, along with either an outer phase shifter ($\phi$) or a tunable directional coupler \cite{sunil_fault_tolerant_programming,Caterina_reconfigurable_quantum_photonic_processor}. The transfer function for a single TBS can be described using the matrix representations for ideal 50:50 beam splitters and lossless phase shifters as: 
\begin{equation} \label{tbs_equation}
\begin{aligned}
T_{BS}&=\frac{1}{2}  
\begin{bmatrix}
     1 & i \\
    i & 1 \\
  \end{bmatrix}
  \begin{bmatrix}
     e^{i\theta} & 0 \\
    0 & 1 \\
  \end{bmatrix}
  \begin{bmatrix}
     1 & i \\
    i & 1 \\
  \end{bmatrix}
  \begin{bmatrix}
     e^{i\phi} & 0 \\
    0 & 1 \\
  \end{bmatrix} \\
  &=ie^{i\frac{\theta}{2}}
  \begin{bmatrix}
     e^{i\phi}sin(\frac{\theta}{2}) & cos(\frac{\theta}{2}) \\
    e^{i\phi}cos(\frac{\theta}{2}) & -sin(\frac{\theta}{2}) \\
  \end{bmatrix}
\end{aligned}
\end{equation}
Thus, any arbitrary light redistribution can be obtained by changing $\theta$ and $\phi$. 

The SVD decomposition of the weight matrix can be described as: 
\begin{equation}
  W=D \prod_{m,n} T_{m,n}
\end{equation}
where $D$ is a diagonal matrix, and $T_{m,n}$ represents the transformation matrix for a $2\times2$ node between two input terminals, $m$ and $n$, within the $N\times N$ multiplication matrix \cite{clements}, given as:

\begin{equation}
T_{m,n}=ie^{i\frac{\theta}{2}}
  \begin{bmatrix}
    1 & 0 & ... &    0    &0 \\
    0 & 1 &  .\quad \quad \quad . &    .    &  0\\
    . & .  & 
    \resizebox{.4\hsize}{!}{$
    \begin{bmatrix}
    e^{i\phi}sin(\frac{\theta}{2}) & cos(\frac{\theta}{2})\\
    e^{i\phi}cos(\frac{\theta}{2}) & -sin(\frac{\theta}{2}) \\
 \end{bmatrix}_{m,n}
 $}
    &  &  \\
    . & . &  .\quad \quad \quad . &     .   &  .\\
    0 & . &   &     1   &  0\\
    0 & 0 & ...  &       0 &  1\\
  \end{bmatrix}
\end{equation}

\subsection{Optical Network Link Budget}

\begin{figure*}[h!]
\centering
\includegraphics[width=1.0\linewidth]{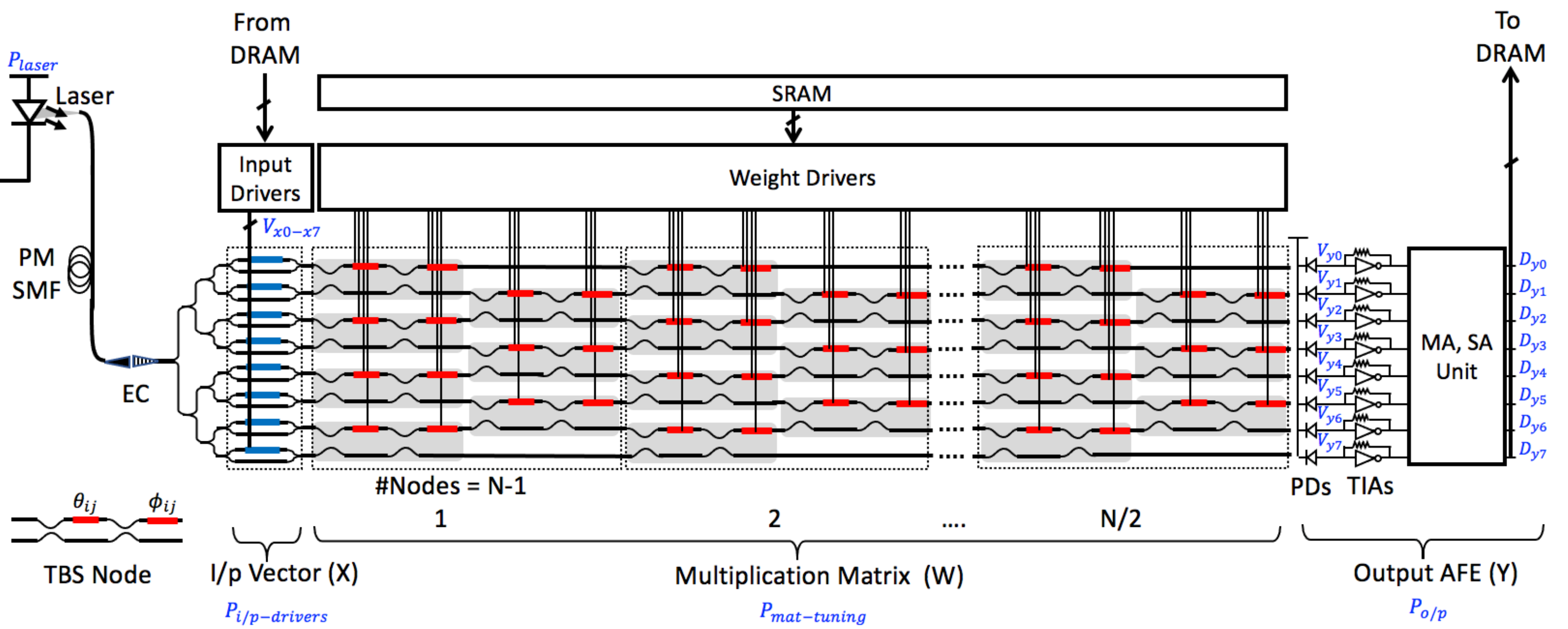}
\caption{SiP circuit diagram of an $N\times N$ MZM-based accelerator. High-speed PN phase shifters and low-speed thermo-optic phase shifters are colored in blue and red, respectively.}
\label{schematic_8N}
\end{figure*}

The photonic components shown in Fig.~\ref{schematic_8N} are simulated in Cadence Spectre for $8\times 8$ and $32\times 32$ transformation matrix sizes to verify the optical link budget analysis. The optical components are modeled in Verilog-A to enable electronics-photonics co-simulation \cite{Sudip_Shekhar_Silicon_Electronic_Photonics, cheryl_veriloga}. The models of some of the components used in this paper are derived from \cite{cheryl_veriloga}, with some modifications to account for the laser electrical power consumption, the wall plug efficiency, $\eta_{WPE}$, link losses, etc. The optical power of laser is set to 0~dBm to estimate the optical power at the output terminals of the network, incident on the PDs. Coupling light to the SiP chip introduces loss in the range of 0.6-to-3~dB based on the coupling scheme used \cite{dietrich_laser_to_sma_loss_0.6dB,honmou_1dB_coupling}. 
The overall coupling loss from the laser to the SiP chip is collectively estimated as 1.6~dB considering possible optimizations in the coupling efficiency. 1.6~dB is also a realistic estimate for photonic wire bonds (PWBs), an emerging technology which involves writing three-dimensional waveguides in a photosensitive polymer. PWBs have demonstrated efficient interfacing between the external sources to the silicon waveguide with coupling losses as low as 0.4~dB up to 1.7~dB \cite{photonic_wirebonding_germany,vanguard_photonic_wirebonding, silicon_photonic_448gbit, nicole_photonic_wire_bond_1.6dB,nicole_photonic_wire_bond_1.7dB}. 

For an input vector size of $N$, light passes through $log_2N$ splitters before modulation, resulting in a total insertion loss of $10log_{10}N+EL_{splitter} \cdot log_2N$ dB, where $EL_{splitter}$ is the estimated excess loss for a single splitter and ranges between 0.01~dB to 0.5~dB \cite{Lukas_Chrostowski_JSTQE2019,yuan_splitter_0.2dB, alireza_pam4_modulator,Daoxin_beamsplitter_0.35db,david_gonzalez_beamsplitter_0.5db}. The estimated excess loss for beam splitters and combiners in this study is 0.01 dB. The overall attenuation in the silicon waveguide is a function of the depth of the network. The MZM based implementation is based on Clement's arrangement which is composed of beam splitters and phase shifters that can be programmed to implement linear transformation\cite{clements}. With an MZM representing one node in Clements arrangement\cite{clements}, the waveguide attenuation can be approximated as $N\eta_{wg}L_{MZI}$, where $\eta_{wg}$ represents the optical intensity attenuation in the Si waveguide and $L_{MZI}$ is the length of an MZM arm, with chosen values of 3~dB/cm and 0.5~mm respectively.  The insertion loss introduced by an MZM's PN phase shifter is approximated as 1~dB/mm \cite{alireza_pam4_modulator}.

The insertion loss through a node is dependent on the excess loss for cross and through states, $EL_{cross}$ and $EL_{thru}$, respectively, both of which are typically $<$1~dB  \cite{Zeqin_Lu_Broadband_Silicon_photonic_directional_couplers, Sitao_Chen_Silicon_directional_coupler}. For simplicity, two phase shifters connected by 3~dB adiabatic directional couplers ($EL_{DC} \sim$ 0.1~dB)\cite{Lukas_Chrostowski_JSTQE2019} are assumed in this work for analysis.

To study the optical attenuation of the $8\times 8$ MZM implementation shown in Fig.~\ref{schematic_8N} and verify its functionality, all inputs, except for the uppermost terminal ($m=1$), are driven by $V_\pi$ voltage that creates a $\pi$ phase shift difference between their MZM's arms and null their outputs.
As light is set to propagate through the uppermost input terminal, the optical depth is defined by the route passing through the diagonal TBS nodes with $i=j$.

For a rectangular mesh arrangement, the matrix optical depth is equal to $N$ with a total number of $\frac{N(N-1)}{2}$ optical crossings \cite{clements,reck}. Hence, the $8\times 8$ matrix implementation shown in Fig.~\ref{schematic_8N}, has an optical depth of 8 with 28 crossings.

The total optical link budgets are calculated based on (\ref{link_budget_equation}), where $P_{SMF-att}$, $P_{EC-IL}$, $P_{Si-att}$, $P_{splitter-IL,EL}$, $P_{PS-IL}$, $P_{DC-IL}$ and  $P_{penalty}$ represent the attenuation introduced by the SMF fiber, fiber to chip coupling loss, silicon waveguide attenuation, splitter insertion and excess loss, phase shifters' insertion loss, total adiabatic coupling insertion loss and network penalty, respectively. The network penalty takes into account further impairments due to extinction ratio, cross talk, intersymbol interference (ISI) and laser relative intensity noise (RIN) which is caused by the random spontaneous emission over time \cite{AR_Silicon_Photonics_MRR_Datacenters,Rongqing_Hui_2020, Can_Li_2016}. 
\begin{equation} \label{link_budget_equation}
\begin{aligned}
  P_{O/p}(dBm) &=P_{laser}-P_{SMF-att}-P_{EC-IL}\\
  &-P_{Si-att}-P_{splitter-IL,EL}\\
  &-N P_{PS-IL}-N P_{DC-IL}-P_{penalty}
\end{aligned} 
\end{equation}

Fig.~\ref{optical_link_budget_N} shows the calculated optical power throughout $N\times N$ networks with different inputs vector sizes. The optical power of the laser is set to $0~dBm$ for ease of illustration. Besides the attenuation due to the splitting, it can be noticed that the losses introduced by the optical components in the multiplication matrix (i.e. directional couplers and phase shifters) pose a limitation for scaling the network due to the highly attenuated optical intensities reaching the outputs. Fig.~\ref{mzm_matrix_power_resolution} shows the optical intensities required at the analog front-end (AFE) to detect a signal with a resolution of $n_{i/p}$ bit. This is obtained by representing the desired output signal and current noises in terms of the received optical intensity as given in Eq.~(\ref{eq_bitresolution_sensitivity}). The blue dotted lines represent the optical power at the matrix outputs and the corresponding bit resolution for a laser intensity of $10~dBm$. It can be shown that the maximum achievable matrix size is $\sim35\times 35$ for binary networks operating at $DR=10~GS/s$. We revisit this calculation again in Section~\ref{Efficiency_Tradeoff_Factors}.


\subsection{Energy Efficiency}
\label{section_MZM_energy_efficiency}
The total electrical power dissipation of the whole network comprises of the power consumed by the laser, input modulators, thermo-optic tuning of the matrix phase-shifters, and the AFE including PDs. Accordingly, for a configuration of size $N \times N$ operating at a data rate of $DR$, the energy efficiency (J/Op) can be calculated as:
\begin{equation} \label{mzm_based_energy_efficiency}
\begin{aligned}
E(J/Op) &= \frac{P_{laser}}{\gamma \cdot 2N^2\cdot DR}+\frac{NP_{i/p-drivers} +2P_{mem-interface}}{2N^2\cdot DR} \\
& +\frac{2(N-1)P_{mat-tuning}+P_{SOA}+P_{o/p-AFE}}{2N\cdot DR} \\
\end{aligned}
\end{equation}
where $P_{laser}$, $P_{i/p-drivers}$, $P_{mem-interface}$, $P_{mat-tuning}$ and $P_{o/p-AFE}$ represent the electrical power dissipated due to the laser, input modulator drivers, data fetch interfacing circuits, matrix tuning and the output AFE circuits, respectively. $P_{SOA}$ represent the electrical power dissipated if a semiconductor optical amplifier (SOA) is used to recover the loss. 

The factor $\gamma$ refers to the energy efficiency enhancement. It can be represented as $\gamma=\rho_{\scalebox{0.5}{opt}}^2\rho_{\scalebox{.5}{SOA}}$ where $\rho_{opt}$ represents the energy scaling due to the loss of precision factor, and will be described later in Section~\ref{Efficiency_Tradeoff_Factors}. $\rho_{\scalebox{.5}{SOA}}$ represents the efficiency enhancement due to an SOA. Assuming an SOA introducing a gain of $\eta_{\scalebox{0.5}{SOA}}$ (in dB), the corresponding enhancement is $\rho_{\scalebox{.5}{SOA}} =  10^{\frac{\eta_{\scalebox{0.5}{SOA}}}{10}}$. The use of an SOA is discussed later in Section IV, but it can be inferred from Eq.~(\ref{mzm_based_energy_efficiency}) that the use of an SOA always degrades the overall energy efficiency.

The amount of power dissipated by the laser is represented in terms of the laser's wall plug efficiency, $\eta_{WPE}$, the optical insertion losses introduced by the SMF fiber, $IL_{SMF}$, the fiber to chip coupling $IL_{EC}$, the silicon waveguide loss, $IL_{WG}$, the input MZM loss, $IL_{i/p-MZM}$, the weight phase shifter loss, $IL_{weight-PS}$, the directional coupler loss, $IL_{DC}$, as well as the receiver's PD sensitivity, $P_{PD-opt}$, as given in Eq.~(\ref{laser_energy_efficiency_equation}).
\begin{equation} \label{laser_energy_efficiency_equation}
\begin{aligned}
P_{\scalebox{.6}{laser}} &= \frac{10^{\frac{IL_{\scalebox{.4}{WG[dB]}}N(L_{\scalebox{.4}{MZI}})}{10}}N    }{IL_{\scalebox{.4}{SMF}}IL_{\scalebox{.4}{EC}}(EL_{\scalebox{.4}{splitter}})^{log_2N}IL_{\scalebox{.4}{i/p-MZM}}(IL_{\scalebox{.4}{PS}})^{2N}} \\
&\times
  \frac{P_{\scalebox{.4}{PD-opt}}}{\eta_{\scalebox{.4}{WPE}}(IL_{\scalebox{.4}{DC}})^{2N}IL_{\scalebox{.4}{penalty}}}
\end{aligned}
\end{equation}

The total length of the waveguide was roughly approximated as the length spanning the optical depth of the matrix only. The output sensitivity is solved based on the targeted bit resolution, $n_{i/p}$ as well as the total noise at the output front-end due to the photodetector shot noise, dark current $I_d$, thermal noise and laser relative intensity noise (RIN) as given in Eq.~(\ref{eq_bitresolution_sensitivity}), with values reported in Table \ref{table_SNR_calculations}. The parameters $R$, $R_L$, $k$ and $T$ represent the PD responsivity, output load resistance, Boltzmann’s constant and the absolute temperature.

\begin{figure*}
\begin{equation}
n_{i/p} = \frac{1}{6.02}\left[20log_{10}\left(\frac{R\times P_{PD-opt}}{\left[\sqrt{2q(RP_{PD-opt}+I_d)+\frac{4kT}{R_L}+R^2P_{PD-opt}^2RIN}+\sqrt{2qI_d+\frac{4kT}{R_L}}\right]\sqrt{DR/\sqrt{2}}} \right)-1.76\right]
\label{eq_bitresolution_sensitivity}
\end{equation}
\end{figure*}



\begin{table}
\caption{SNR Calculation Parameters} 
\centering  
\begin{ruledtabular}
\begin{tabular}{c c c}  
\quad\quad $ Parameter$ \quad\quad & $Description$&\quad\quad $Value$ \quad\quad  \\ [0.5ex]  
\hline 
$P_{laser}$     &  $Laser~Power~Intensity$ & $10~dBm$ \\
$R$             &  $PD~responsivity$ & $1~A/W$ \cite{gf2019}\\
$R_L$           &  $Load~Resistance$ & $50~\Omega$ \\
$I_d$           &  $Dark~Current$ & $35~nA$ \cite{gf2019}\\
$T$             &  $Absolute~Temperature$ & $300~K$ \\
$DR$       &  $Data~Rate$ & $10~GS/s$ \\
$B_o$           &  $Optical~Bandwidth$ & $25~GHz$ \\
$B_e$           &  $Electrical~Bandwidth$ & $DR/\sqrt{2}~GHz$ \\
$\lambda$       &  $Wavelength$ & $1550~nm$ \\
$RIN$           & $Relative~Intensity~Noise$  & $-140~dB/Hz$ \cite{Rongqing_Hui_2020, Can_Li_2016}\\
$WPE$      & $Wall~Plug~Efficiency$  & $10\%$ \\
\end{tabular}
\end{ruledtabular}
\label{table_SNR_calculations}  
\end{table}

The MZM drivers consume power that scales linearly with $N$ as $P_{mod-driver}=N\cdot DR \cdot E_{MZM-driver}$, where $E_{MZM-driver}$ represents the energy efficiency of the MZM driver. For binary resolution, the power consumed by the drivers and AFEs are extracted from recent work on PAM2, and is typically in the range of $\sim$2pJ/b
 \cite{Laszio_Szilagyi_mach_zhender_modulator_driver}. 

The matrix weights are tuned using thermo-optic phase shifters (TO-PS). Doped Si heaters on SOI platform typically dissipate about $\sim$ 20~mW for a $\pi$-shift \cite{AlTaha_Monitoring_and_automatic_tuning_and_stabilization_of_2x2_mzi_optical_switch, Maxime_Jacques_Optimization_of_thermo_optic_phase_shifter_design, Nicholas_Harris_efficient_compact_low_loss_thermo_optic_phase_shifter}. The efficiency of TO-PS can be improved using other heater materials such as TiN, substrate undercut to improve insulation and deep trenches to reduce thermal cross-talk \cite{Maxime_Jacques_Optimization_of_thermo_optic_phase_shifter_design, Shigeru_Nakamura_high_extinction_ratio_optical_switching}. This can be shown to significantly improve the overall energy efficiency of the network as illustrated in Fig.~\ref{energy_efficiency_rho1}. Assuming uniformly distributed weights, the expected energy consumption of the thermo-optic phase shifter is $E_{TO-PS} =\frac{1}{P{_\pi}} \int_{0}^{P_{\pi}}P_{heater} dP_{heater} = \frac{P_{\pi}}{2}$, where $P_{\pi}$ denotes the amount of electrical power required to create a phase shift of $\pi$. Calculating for all the nodes in Clement's topology, the total average tuning power is $\frac{N(N-1)}{4}P_{\pi}$.

After optical processing, the optical data needs to be converted to the electrical domain to be processed, stored or reused in other networks. Efficient opto-electronic receivers, comprising of a PD, TIA and main amplifiers have been shown to have energy efficiencies $\sim0.4 -2.4~$pJ/b \cite{Frankie_Liu_Optical_Transmitter_and_Receiver_Circuits_in_40_nm_CMOS, Yuan_Sheng_Lee_Fully_Integrated_Optical_Receiver, Takashi_Takemoto_transimpedance_amplifier_in_65nm_CMOS, Hiroshi_Morita_two_dimentional_optical_IO_array_chip_to_chip_interconnect_65nm, Clint_Schow_A_Single_Chip_CMOS_Based_Parallel_Optical_Transceiver_Bidirectional_Data_Rates, Kyu_Sang_Park_optical_receiver_front_end_with_5_mW_transimpedance_amplifier}. For an AFE operating at $10~$Gb/s and realized in $40~nm$ CMOS technology, an energy efficiency of $0.4$~pJ/b\cite{Frankie_Liu_Optical_Transmitter_and_Receiver_Circuits_in_40_nm_CMOS} is assumed for the calculation of binary resolution AFE, which scales with a factor of $N$ for the whole output array. Higher AFE resolutions entail the use of linear TIAs along with analog to digital converter (ADC) circuits to recover the digital data. High speed linear TIAs have shown efficiencies as low as $0.6$~pJ/b \cite{Linear_TIA_Lakshmikumar2019}. The energy consumption for ADCs is extracted from the energy per conversion figure of merit (FOM) such that $E_{ADC}~ (J/b)=2^N\times FOM$. The energy consumption values used in this work for $2b$, $3b$ and $4b$ are $1.7$~pJ/b, $3.1$~pJ/b and $5.7$~pJ/b based on a FOM of $0.335~$pJ/conversion for an ADC designed to operate at a sampling rate of $28~$GS/s \cite{56Gbps_PAM4_TI_SAR_Frans2017}.

Providing high-speed serial inputs to the SiP accelerator requires FIFOs and multiplexers to interface the data transfer with DRAM, as shown in Fig.~\ref{fig_data_fetch}. For a fair comparison to digital CMOS implementations, the power dissipation for both input and output interfacing circuits, represented by $P_{mem-interface}$ is taken into consideration in the energy efficiency calculation of SiP implementations. The power dissipated by the FIFO, multiplexers, clock dividers and retimers is estimated as $5.77~mW$ in $28~nm$ CMOS based on the data reported in $\cite{Green_2002}$ in $180~nm$ CMOS technology. 

\subsection{Efficiency Tradeoff Factors}
\label{Efficiency_Tradeoff_Factors}

\begin{figure*}
\centering\includegraphics[width=0.710\linewidth]{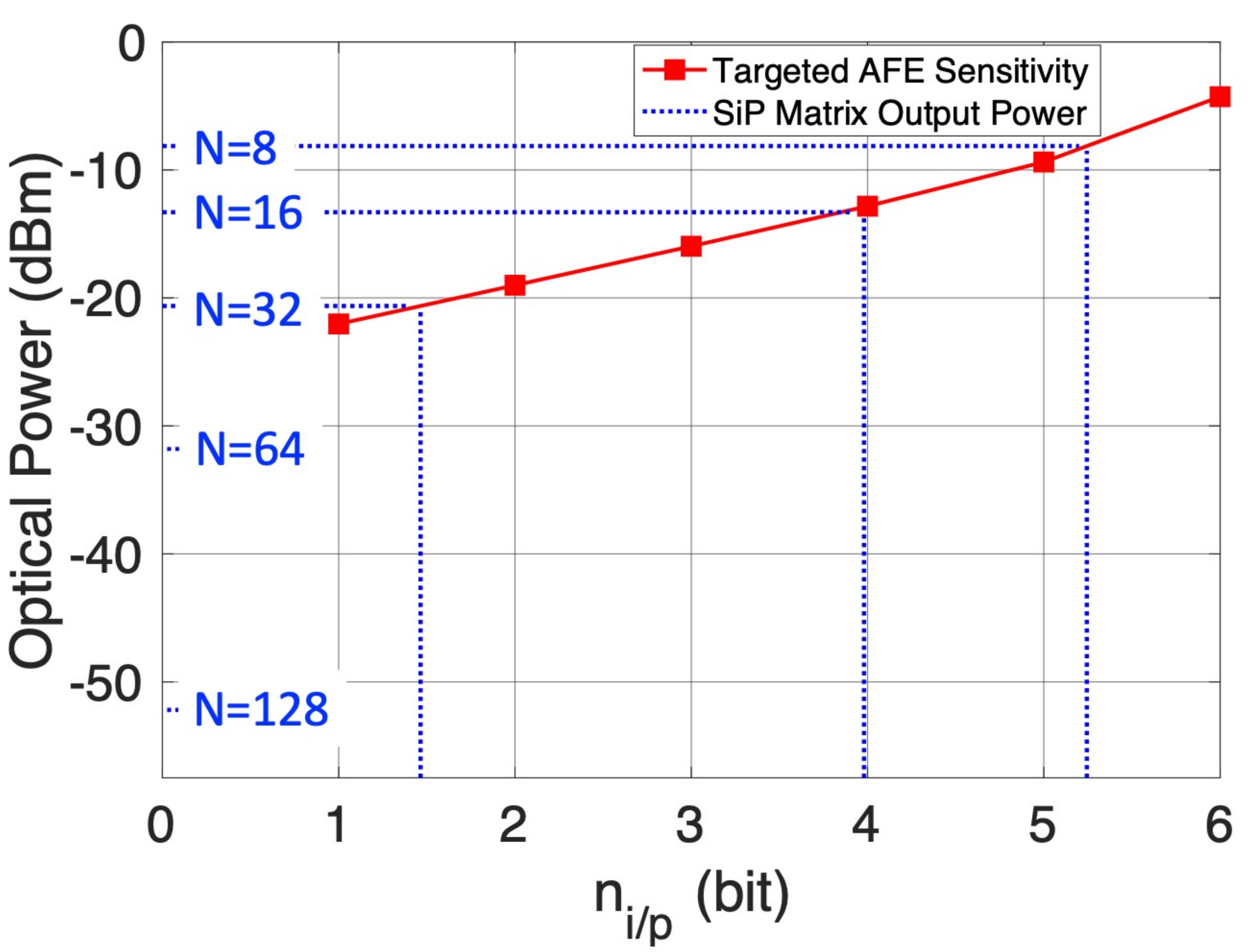}
\caption{Targeted AFE sensitivity for $n_{i/p}=\{1,2,3,4,5,6\}b$ and the output power for MZM-based SiP matrices with sizes N = 8, 16, 32, 64, 128. $P_{laser} = 10~dBm$.}
\label{mzm_matrix_power_resolution}
\end{figure*}

\begin{figure*}[h!]
\centering\includegraphics[width=0.79\linewidth]{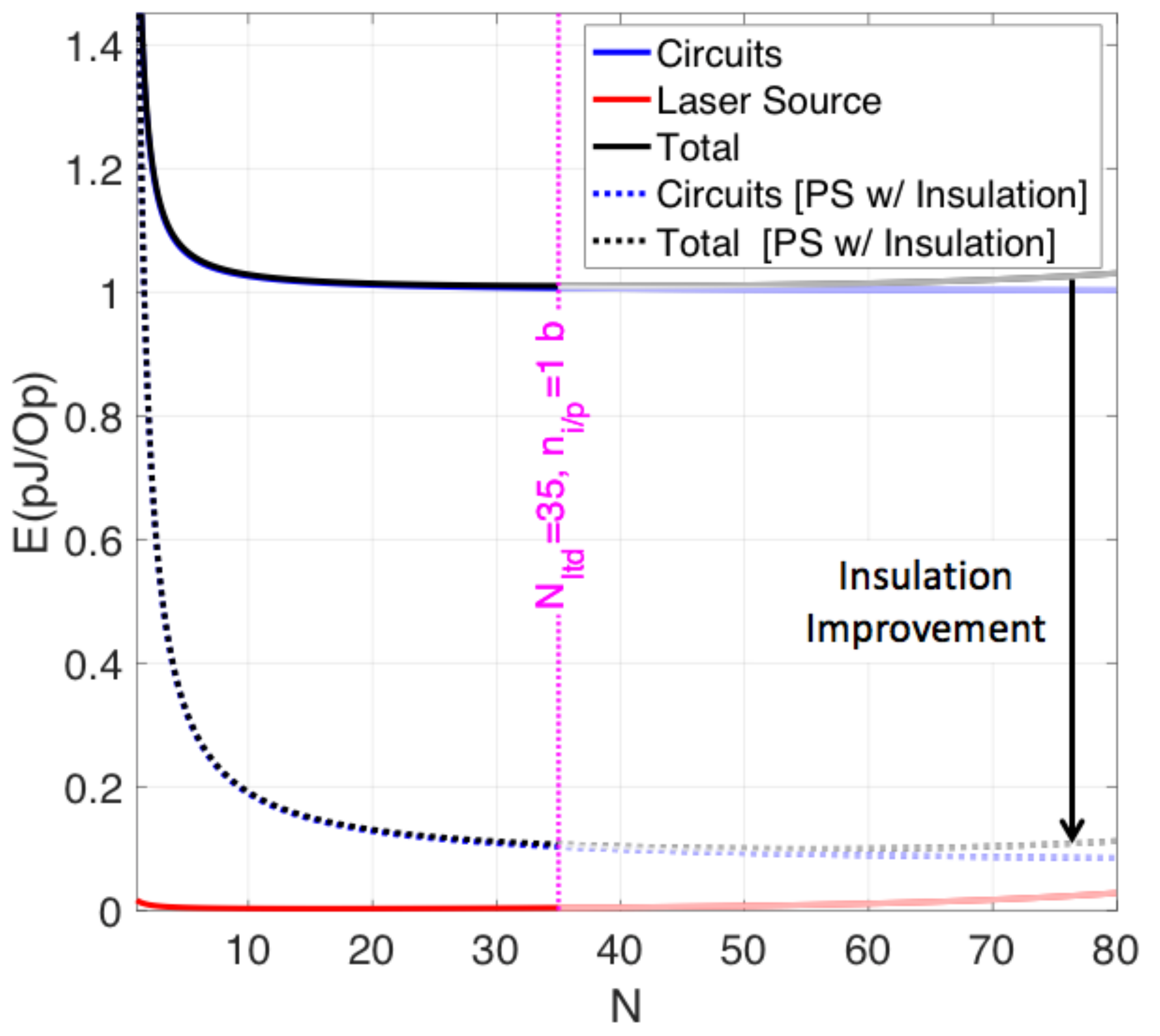}
\caption{Total energy efficiency (pJ/Op) for an $N\times N$ MZM implementation with PD responsivity $R=1.2~A/W$ and binary resolution, $n_{i/p}=1b$. Energy efficiency improves as the network scales up due to the increased number of operations. Improving the tuning efficiency with insulation has significant enhancement on the overall energy efficiency.}
\label{energy_efficiency_rho1}
\end{figure*}

It can be inferred from Eq.~(\ref{laser_energy_efficiency_equation}) that the required input laser power increases as a function of $N$ as a result of the exponentially increasing optical losses in the MZM based accelerator. But the total energy efficiency Eq.~(\ref{mzm_based_energy_efficiency}) starts improving as $N$ scales up due to the quadratic increase in the number of accelerated operations performed by the optical matrix as shown in Fig.~\ref{energy_efficiency_rho1}. Taking all optical losses into account shows that there is a scaling limit beyond which optical losses grow significantly and the overall efficiency drops and an optimal network size exists for minimum energy efficiency. Unfortunately, the maximum network scale, $N_{ltd}$, is limited by the rated output optical power of the laser and the SNR required for any given signal resolution, $n_{i/p}\geq1b$, as given in Eq.~(\ref{eq_bitresolution_sensitivity}) and illustrated in Fig.~\ref{mzm_matrix_power_resolution}. 

\begin{figure*}
\centering\includegraphics[width=0.810\linewidth]{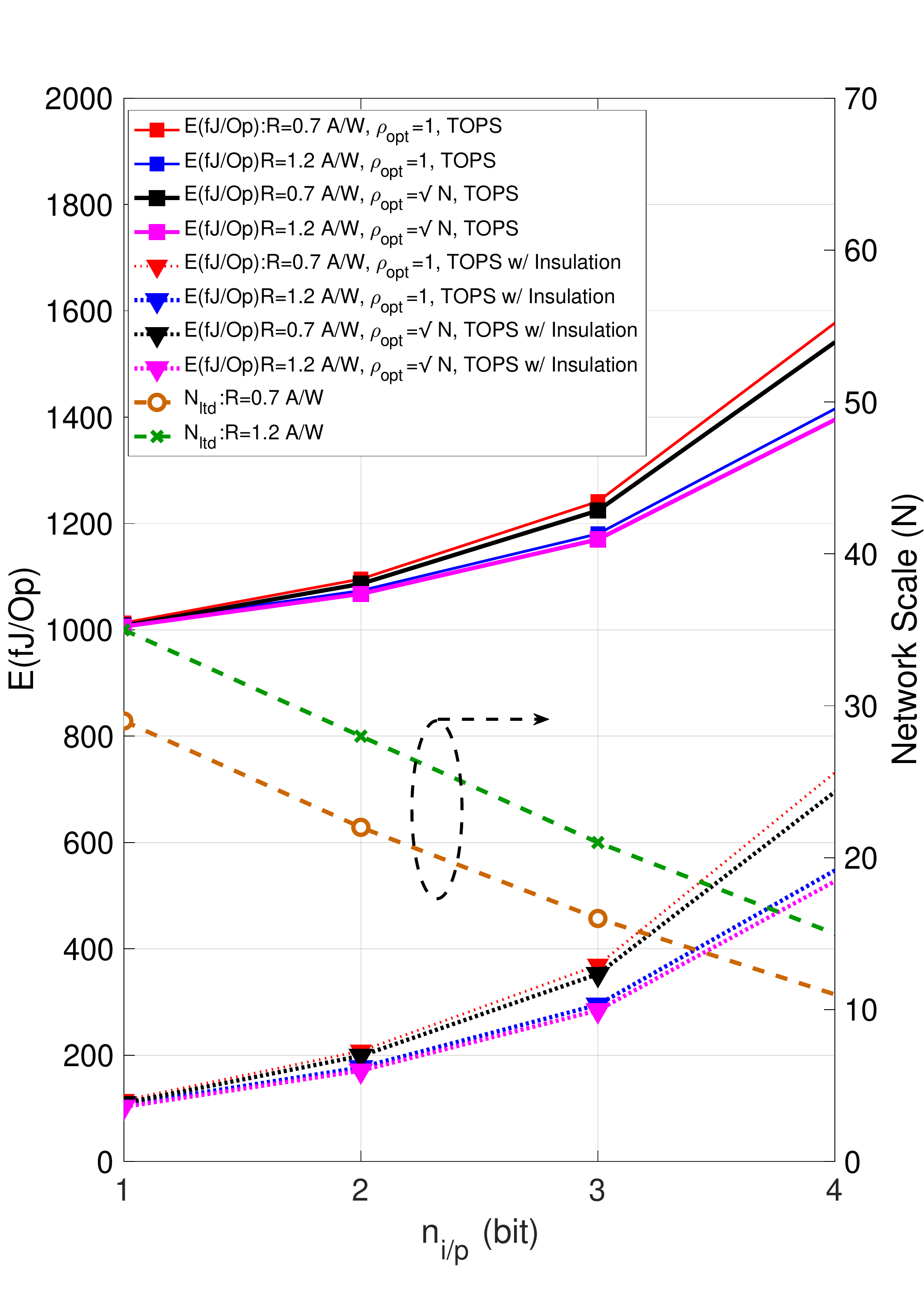}
\caption{Total energy efficiency and scaling limit of an MZM based network with an input resolution of $n_{i/p} = \{1,2,3,4\}b$ considering thermo-optic phase shifters with and without insulation at the scaling limit, $N_{ltd}$, where the laser rated power output (10~dBm) is reached. Implementing networks with higher resolution requires scaling down the network to improve the SNR at the AFE. This leads to a degradation of the energy efficiency due to the reduced number of operations.}
\label{energy_efficiency_sensitivity_precision}
\end{figure*}

Fig.~\ref{energy_efficiency_sensitivity_precision} shows the total energy efficiency and scaling limit for various input resolutions considering thermo-optic phase shifters with and without insulation. The energy efficiencies in Fig.~\ref{energy_efficiency_sensitivity_precision} are calculated for accelerators to be operated at binary and higher resolution, $n_{i/p}=\{1,2,3,4\}b$. Although the probability of transition reduces for multilevel signaling \cite{Sajjad_moazeni_Vladimir_A_40Gbps_PAM4_Tx_based_on_a_ring_resonator_optical_DAC_in45nm_SOI_CMOS}, the requirement on driver's linearity or segmentation also increases. Furthermore, the energy consumed in the serializing and clocking remains the same \cite{Sajjad_moazeni_Vladimir_A_40Gbps_PAM4_Tx_based_on_a_ring_resonator_optical_DAC_in45nm_SOI_CMOS}. Thus, we assume similar energy efficiency for multilevel signaling as PAM2, $\sim2pJ/b$, for MZM modulators\cite{56Gbps_PAM4_Tanaka2018}. For binary resolution, the power consumed by the drivers and AFEs are extracted from recent work on PAM2 transceivers\cite{56Gbps_PAM4_Tanaka2018, Sajjad_moazeni_Vladimir_A_40Gbps_PAM4_Tx_based_on_a_ring_resonator_optical_DAC_in45nm_SOI_CMOS, MZM_Tx_20Gbps_Michard2020, Linear_TIA_Lakshmikumar2019,56Gbps_PAM4_Tanaka2018}. Therefore, the energy efficiency for $2b$, $3b$ and $4b$ input MZM drivers are estimated in our calculation as $\sim4pJ$, $6pJ$ and $8pJ$ per symbol, respectively. 

It can be concluded from Fig.~\ref{energy_efficiency_sensitivity_precision} that opting for PDs with higher responsivities improves the energy efficiency of the network. This compensates for the optical system loss, relaxes the need to inject high optical power at the network input and improves the overall energy efficiency. Utilizing avalanche PDs (APDs) is a possible way to significantly improve the optical sensitivity \cite{Spoorthi_Nayak_10Gbps_18.8dBm_Fully_Integrated_Optoelectronic_Receiver_With_Avalanche_Photodetector}. Fig.~\ref{energy_efficiency_sensitivity_precision} also suggests that taking advantage of the loss of precision, when possible, shows minor improvement in the energy efficiency and the network scaling. 

\begin{table}
\caption{MZM-based Implementation Characteristics} 
\resizebox{\columnwidth}{!}{
\centering 
\begin{ruledtabular}
\begin{tabular}{c c c c c c c c} 
$\eta_{\scalebox{0.6}{WPE}}$ &$IL_{\scalebox{0.6}{SMF}}$&$IL_{\scalebox{0.6}{EC}}$&$IL_{\scalebox{0.6}{WG}}$&$EL_{\scalebox{0.6}{Splitter}}$&$IL_{\scalebox{0.6}{MZI}}$&$L_{\scalebox{0.6}{MZI}}$&$IL_{\scalebox{0.6}{DC}}$  \\ [0.5ex] 
 &$[dB]$&$[dB]$&$[dB/mm]$&$[dB]$&$[dB/mm]$&$[mm]$&$[dB]$ \\ 
\hline
$0.1$&$0$&$1.6$&$0.3$&$0.01$&$1$&$0.5$&$0.01$\\
\end{tabular}
\end{ruledtabular}
\label{MZM_characteristics_table} 
}
\end{table}


Considering a matrix that scales with $N$, the laser optical intensity should be typically scaled by a factor of $N$ to account for the splitting loss in a lossless network. For mesh-like configurations similar to Fig.~\ref{schematic_8N}, scaling the input vector size increases the dynamic range of the output intensities. In other words, for a given matrix output, the intensity can be as low as that of a single input or as high as $N$ times that amount. With the input's digital resolution being $n_{i/p}$, the effective overall output resolution due to the network scaling is $n_{i/p}+log_2N$. 

\begin{figure*}
\centering
\includegraphics[width=0.810\linewidth]{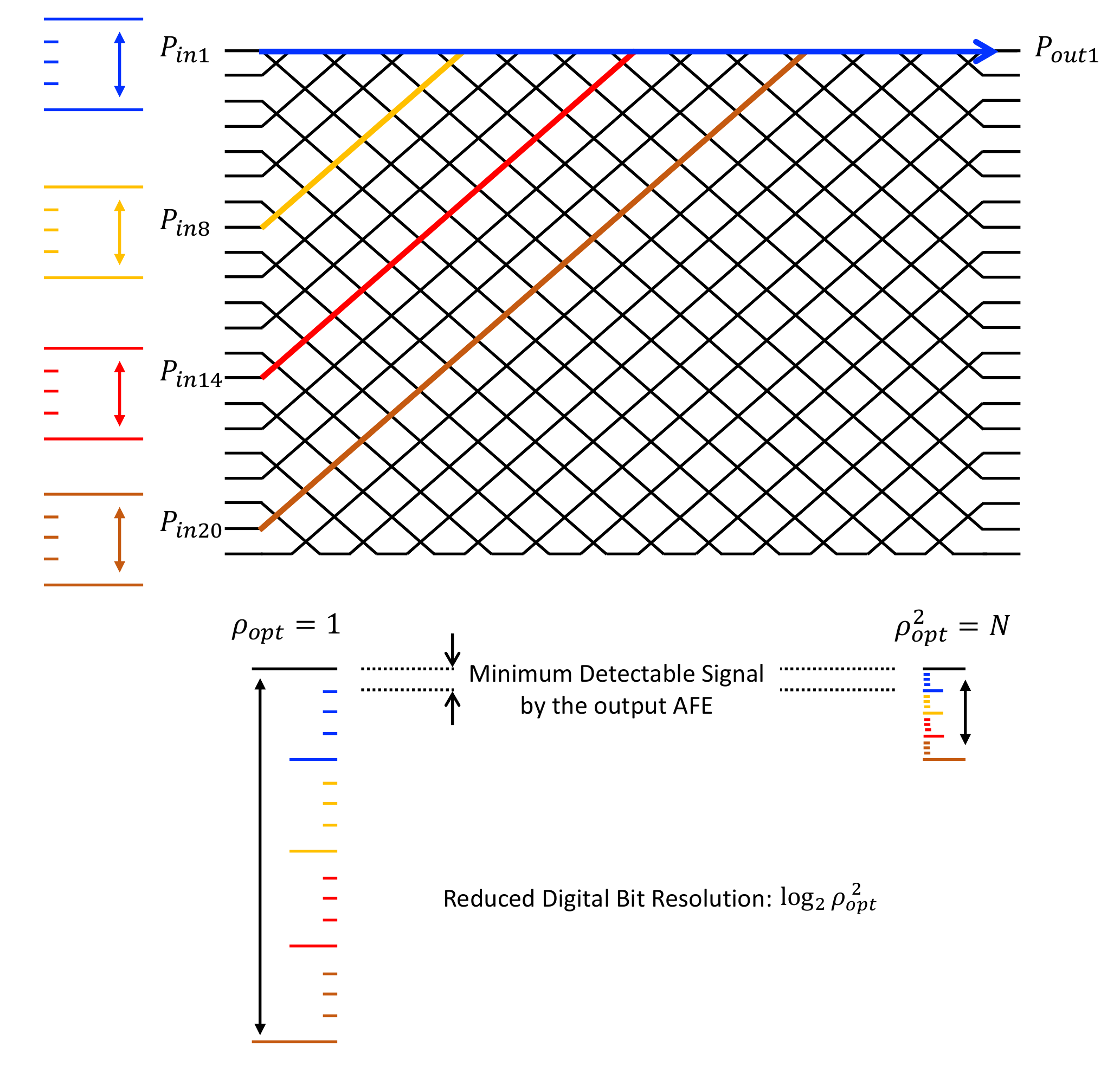}
\caption{The loss of precision factor for a case when $4$ input signals ($2b$ digital resolution) combine to form the output $P_{out1}$. Scaling down the input optical power by $N=4$ can be achieved at the expense of reducing the digital precision at the output by $log_{2}4=2b$.}
\label{loss_of_precision_figure}
\end{figure*}

The conservative estimate of scaling the input power by $N$ may not be necessary in some computational context such as convolutional neural network (CNN) layers with adaptable hidden layer resolutions \cite{Bert_Moons_et_al_Energy_efficient_ConvNets_through_approximate_computing}; the increased output resolution might be higher than that needed by the AFE to detect. Therefore, an energy scaling vs. loss of precision trade-off factor, $\rho_{opt}$, can be introduced to take advantage of the network scaling, \cite{Mitchell_Nahmias_Photonic_Multiply_Accumulate_Operations_for_Neural_Networks,Sapan_Agarwal_Energy_Scaling_Advantages_of_Resistive_Memory_Crossbar_Based_Computation_and_Its_Application_to_Sparse_Coding}, as illustrated in Fig.~\ref{loss_of_precision_figure}. Full accuracy is described by $\rho_{opt} =1$ corresponding to reduced output precision of $log_2(\rho_{opt})=0$, at which the input optical intensity is scaled by $N$. Generally, for $log_2(\rho_{opt})$ bit reduction, the input is scaled by $N/\rho_{opt}$. Therefore, the maximum amount of energy saving is achieved when the $log_2N$ bit reduction is tolerable at the optical output (AFE input). 



To get a meaningful sense of the trade-off between the energy scaling and the loss of precision, $\rho_{\scalebox{.5}{opt}}$ is quantified in Eq.~(\ref{rho_q_factor}) in terms of the probability of bit errors for binary networks (networks with binary weights) at the output such that:
\begin{equation} \label{rho_q_factor}
  Prob_{error} = Q\left( \frac{P_{opt-o/p}}{\rho_{\scalebox{.5}{opt}} }\frac{R}{2i_{irn}}\right)
\end{equation}
where the $Q$ function is defined as $Q(x)=\int\limits_x^\infty \frac{1}{\sqrt{2\pi}} e^{-u^2/2}du$, and $i_{irn}$ represents the total input referred noise at the AFE input with contributions from the PD, TIA, main amplifiers and comparators (if applicable). Therefore, the laser power, $P_{laser}=NP_{opt-o/p}$, can be traded off for loss of output resolution.

The IL difference between the bar and cross states of a tunable beam splitter impacts the interference between the nodes in the mesh. For an MZM with intensity loss of $\alpha_1$ in one arm and $\alpha_2$ in the other, it can be shown that the output intensity at the cross state is given by $I_{cross}=I_{in1}[\alpha_1+\alpha_2+2\sqrt{\alpha_1\alpha_2}cos(\theta_2)]$ when $I_{in2=0}$, where $I_{in1}$ and $I_{in2}$ represent the MZM input intensities at its input ports 1 and 2, respectively. To get the transmission response of an MZM with equal losses, $\alpha_1$ on both arms, $\theta$ should be modified such that $cos(\theta)=\frac{\Delta\alpha + 2\alpha_1cos(\theta_1)}{2\sqrt{\alpha_1\alpha_2}}$ in order to account for the loss difference between the two MZM arms, where $\Delta\alpha=\alpha_1-\alpha_2$ and $\theta_1$ describes the phase shift when both arms have attenuation of $\alpha_1$. This difference in insertion losses can be observed in single-arm beam splitters in which phase shifters are controlled by a single arm only. Using dual-arm tunable beam splitters, where $\theta$ is implemented differentially using phase shifters on both arms, introduces equal insertion losses for the bar and cross transmissions of each node. A dummy phase shifter can also be used in single-arm topologies to obtain equal losses.

For sake of comparison, the energy consumption for a digital MAC is estimated based on the energy consumed by multiplication and accumulation operations as well as register file access in a $28~nm$ CMOS implementation
\cite{Gudaparthi_Sumanth_Wire_Aware_Architecture_and_Dataflow_for_CNN_Accelerators}. With an estimated energy consumption 0.046~pJ for 8b MAC and 0.0117~pJ for register file access, the calculated energy consumption is $\sim (0.046~$pJ$ + 0.0117~ $pJ$)/2 $=$ 28.85~$fJ for a single operation. Conversely, it can be observed from Fig.~\ref{energy_efficiency_sensitivity_precision} that SiP networks based on MZMs need to be scaled down to achieve higher resolutions which further degrades their energy efficiency. Compared to their 8b digital CMOS counterpart, the energy efficiencies for SiP MZM MACs (using low-power thermo-optic phase shifters with insulation and 1.2~A/W\cite{Daniel_Benedikovic_Si_Ge_PD_1.2A_W} PD responsivity) at $n_{i/p}$ = 1b to 4b resolutions are $~3.5\times$ to $~17.5\times$  worse. Despite the lower energy efficiency, MZM-based MAC operations are performed at 2$N\times$ higher operating speed and lower latency (for a weight-stationary systolic array with input vector size of $N\times 1$) than the corresponding digital CMOS implementation. 

In addition, multiple clock cycles are needed for digital multipliers and adders to provide the MAC output in systolic arrays. Operating at $\sim 10\times$ lower clock speeds further decreases their throughput in comparison to optical implementations. Assuming a number $\alpha$ of clock cycles needed for digital MAC, the latency of a digital systolic array with a $1\times N$ input vector size and a $N\times N$ matrix size is $2N\alpha/f_{CLK-CMOS}$ where $f_{CLK-CMOS}$ represent the clock frequency of the digital CMOS implementation. Hence, the throughput ratio of the optical to CMOS implementations is $2N\alpha f_{CLK-OPT}/f_{CLK-CMOS}$, where $f_{CLK-OPT}$ represents the clock speed at which an optical implementation operates at.

\begin{figure*}
\centering\includegraphics[width=1.0595\linewidth]{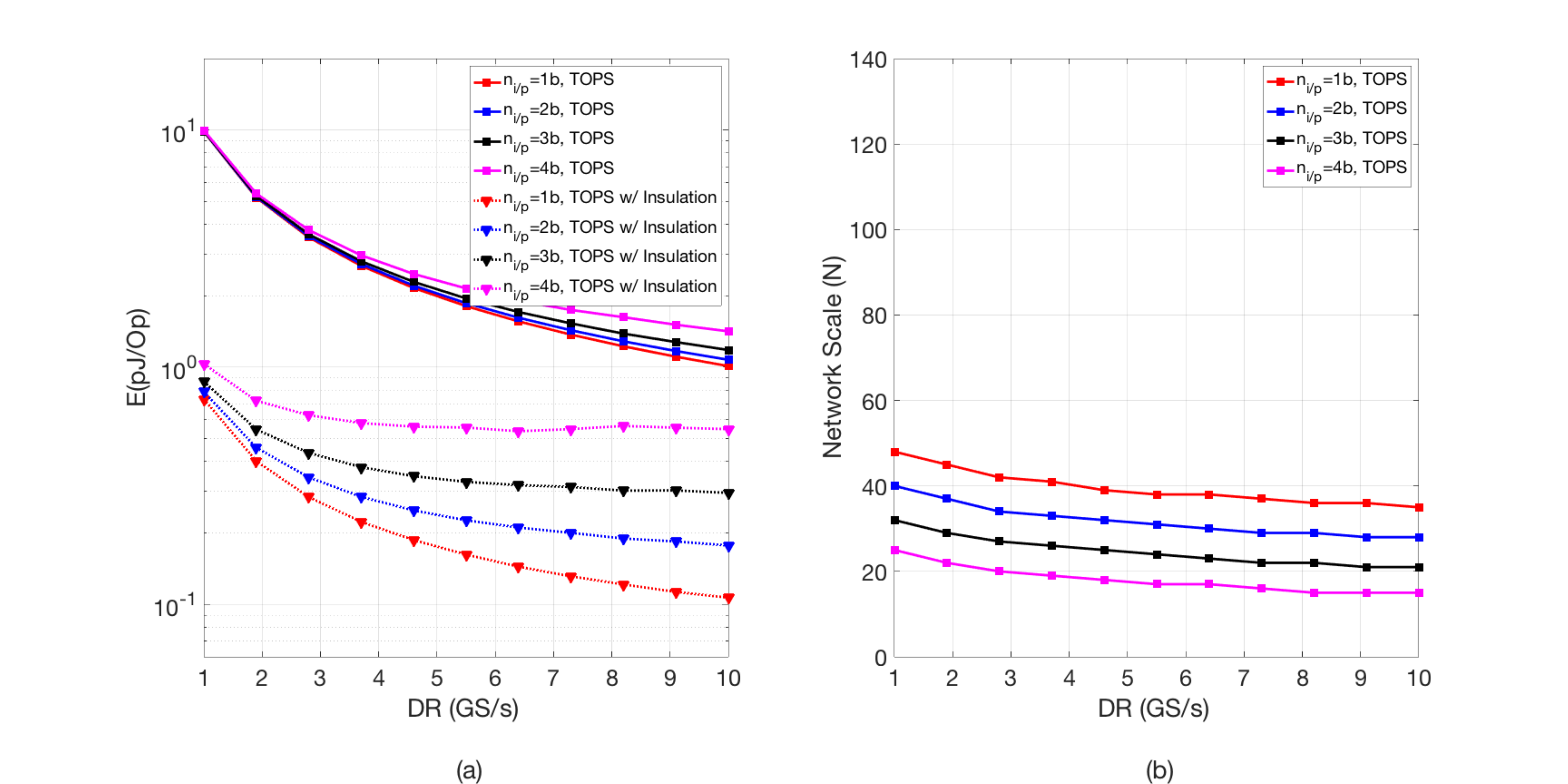}
\caption{(a) Total energy efficiency and (b) MZM Network sizes for $R=1.2~A/W$, $DR=1-10~GS/s$. Operating at higher data rates helps reduce the contribution of the static power consumption of thermo-optic phase shifters to the energy efficiency. Network scales up at lower data rates due to the SNR improvement.}
\label{fig_E_N_vs_DR}
\end{figure*}

We also investigate the energy efficiency and network size at lower data rates in Fig.~\ref{fig_E_N_vs_DR}. Intuitively, the energy efficiency degrades at lower data rates because less number of operations are conducted with respect to the dissipated static power. On the other hand, the network size, shown in Fig.~\ref{fig_E_N_vs_DR}, can be increased by making use of the SNR improvement at lower data rates, as inferred from Eq.~(\ref{eq_bitresolution_sensitivity}). If maximizing the optical throughput is not an overarching goal, opting for lower data rates (relative to 10~GS/s) leads to larger networks while not sacrificing much on the energy efficiency in implementations incorporating phase shifters with insulation (which do not consume much static power as shown by the dotted curves in Fig.~\ref{fig_E_N_vs_DR}~(a)).

\section{MRR Based Si-photonic Implementation}
\label{Section_MRR_based_SiP_Implementation}

\subsection{System Architecture}
\begin{figure*}
\centering\includegraphics[width=1.07\linewidth]{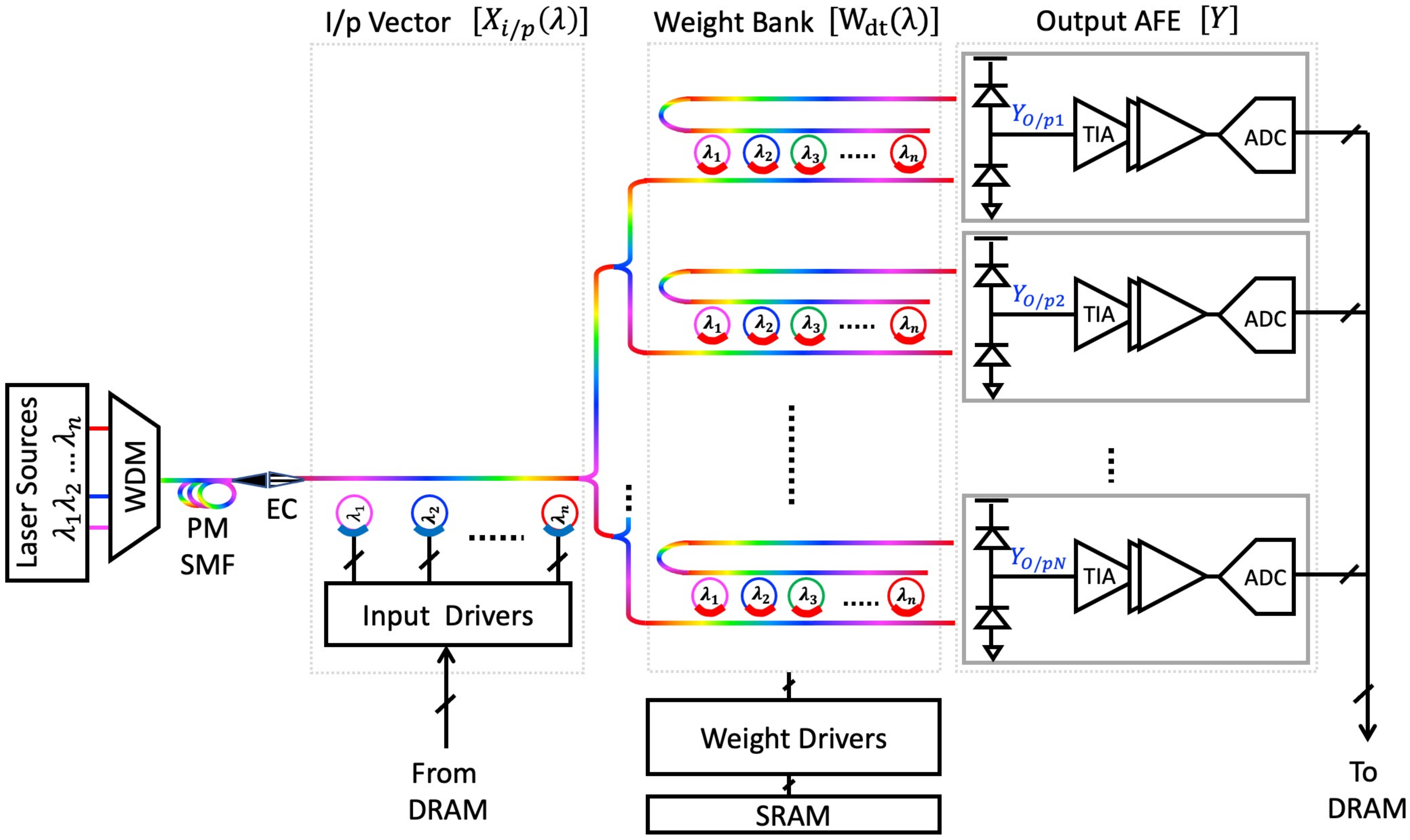}
\caption{SiP circuit diagram of an $N\times N$ MRR-based accelerator}
\label{_mrr_circuit_implementation}
\end{figure*}

The SiP MRR-based implementation of an optical accelerator is illustrated in Fig.~\ref{_mrr_circuit_implementation}. 
A comb CW laser source is used to provide wavelengths $\lambda_1$ through $\lambda_n$ which are coupled into the chip and then modulated by an array of $N$ MRMs. Unlike mesh-like topologies, implementing vector matrix multiplication in the form of dot products has the advantage of maintaining equal path loss for all the outputs. The modulated input vector, $X_{i/p}(\lambda)$, is then split (broadcast) into $N$ branches to be modulated by the weight bank arrays \cite{Tait2014}; each output represents the dot product of the input vector and one of the row arrays of the weight matrix. In order to achieve weights with positive and negative polarities, the thru and drop transmissions of the weight arrays are routed to balanced PDs in a push-pull configuration at the receiver. The current difference at the output, $Y_{O/p}$, can be represented as \cite{Viraj_Bangari_Digiral_Electronics_and_Analog_Photonics_DEAP_CNN}:
\begin{equation}\label{average_pd_current_rx}
\begin{aligned}
    Y_{O/p} = \int_{-\infty}^{\infty} |E_0(\lambda)|^2 X_{i/p}(\lambda)W_{dt}(\lambda) R(\lambda) d\lambda
\end{aligned}
\end{equation}
where $E_{0}(\lambda)$, $W_{dt}(\lambda)$ and $R(\lambda)$  represent the amplitude of the input optical field, the difference between the weight's drop and thru intensity transmissions and the PD responsivity at a wavelength $\lambda$, respectively.

\subsection{Optical Network Link Budget}
Similar to Eq.~(\ref{link_budget_equation}), the optical link budget is calculated based on Eq.~(\ref{mrr_link_budget_equation}): 
\begin{equation} 
\resizebox{.95\hsize}{!}{
\label{mrr_link_budget_equation}
$
\begin{aligned}
P_{O/p}(dBm) &=P_{laser}-P_{SMF-att}-P_{EC-IL}-P_{Si-att} \\
  &- P_{MRM-I/p-IL}- (N-1) P_{MRM-I/p-OBL} \\
  &-P_{splitter-IL,EL} - P_{MRR-W-IL}\\
  &- (N-1) P_{MRR-W-OBL}-P_{penalty}  
\end{aligned}
$
}
\end{equation}
where $P_{MRM-I/p-IL}$ represents the transmission insertion loss of the MRM for the input vector, $P_{MRM-I/p-OBL}$ represents out of band insertion loss ($OBL$) of the MRM for the input vector when the MRM resonance wavelength does not match the input vector wavelength, $P_{MRR-W-IL}$ represents transmission insertion loss of the MRR for the weight vector, and $P_{MRR-W-OBL}$ represents out of band insertion loss of the MRR for the weight vector. Other terms have been defined already when describing Eq.~(\ref{link_budget_equation}). 


\begin{table}
\caption{MRR-based Implementation Characteristics} 
\centering  
\begin{ruledtabular} 
\begin{tabular}{c c  }  
\quad\quad $ Parameter$ \quad\quad & \quad\quad $Value$ \quad\quad  \\ [0.5ex]  \hline 
$\eta_{\scalebox{0.6}{WPE}}$    &    $0.1$ \\
$IL_{\scalebox{0.6}{SMF}}[dB]$    &    $0$ \\
$IL_{\scalebox{0.6}{EC}}[dB]$    &    $1.6$ \\
$IL_{\scalebox{0.6}{WG}}[dB/mm]$    &    $0.3$ \\
$EL_{\scalebox{0.6}{Splitter}}[dB]$    &    $0.01$ \\
$IL_{\scalebox{0.6}{MRM}}[dB]$\cite{Hao_Li_PAM4_TX_JSSC2021}  &    $4$ \\
$OBL_{\scalebox{0.6}{MRM}}[dB]$  &    $0.01$ \\
$IL_{\scalebox{0.6}{MRR}}[dB]$  &    $0.01$ \\
$d_{\scalebox{0.6}{MRR}}[\mu m]$  &    $20$ \\
$IL_{\scalebox{0.6}{penalty}}[dB]$  &    $4.8$ \\
\end{tabular}
\end{ruledtabular} 
\label{MRR_characteristics_table}  
\end{table}

\begin{figure*}
\centering\includegraphics[width=0.710\linewidth]{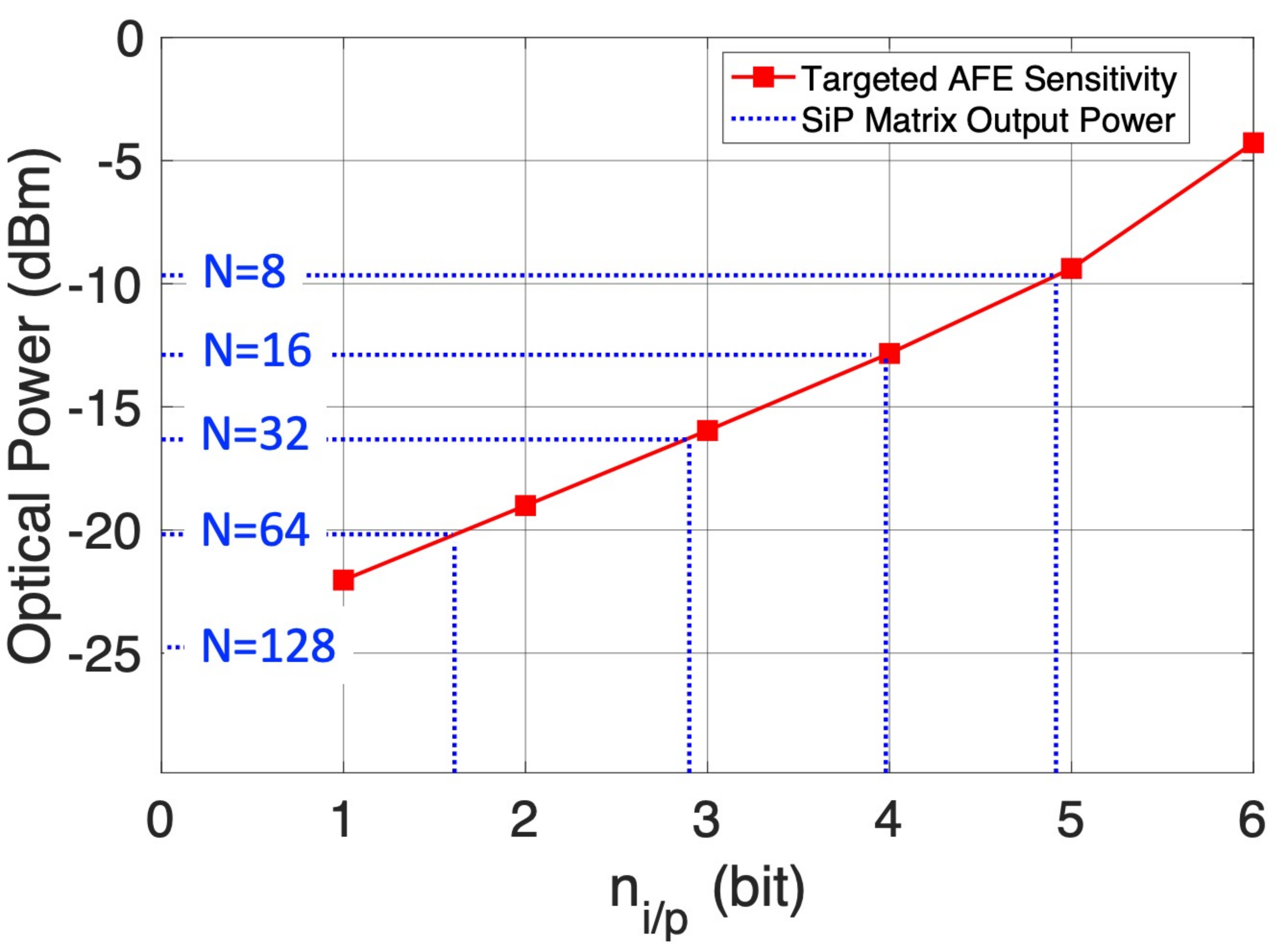}
\caption{Targeted AFE sensitivity for $n_{i/p}=\{1,2,3,4,5,6\}b$ and the output power for MRR-based SiP matrices with sizes N = 8, 16, 32, 64, 128. $P_{laser} = 10~dBm$.}
\label{mrr_matrix_power_resolution}
\end{figure*}

Fig.~\ref{optical_link_budget_N_mrr} shows the calculated optical power throughout an MRR-based implementation with different input vector sizes, $N$, using the values shown in Table~\ref{MRR_characteristics_table}. The optical power of the laser is set to $0~dBm$ for the ease of illustration. Similar to MZM-based implementations, the attenuation introduced due to the splitting and cascading of microrings significantly degrade the optical power and pose a limitation on the energy efficiency as will be discussed in Section \ref{MRR_Energy_Efficiency}. Fig.~\ref{mrr_matrix_power_resolution} shows the optical intensities required at the AFE to detect a signal with a resolution of $n_{i/p}$ bit. This is obtained by representing the desired output signal and current noises in terms of the received optical intensity as given in Eq.~(\ref{eq_bitresolution_sensitivity}). It can be shown that the maximum achievable matrix size is $\sim 85\times 85$ for binary networks. We revisit this calculation again in Section~\ref{MRR_Tradeoffs}.

\subsection{Energy Efficiency} \label{MRR_Energy_Efficiency}

The energy efficiency $(J/Op)$ of the MRM-based implementation with size $N\times N$ operating at a data rate of $DR$ can be calculated as:
\begin{equation} \label{mrr_based_energy_efficiency}
\begin{aligned}
E(J/Op)&= \frac{P_{laser}}{\rho_{\scalebox{.5}{SOA}}\cdot2N^2\cdot DR}  + \frac{NP_{i/p-drivers} +2P_{mem-interface}}{2N^2\cdot DR}\\
&  + \frac{N P_{mat-tuning}+P_{SOA} +P_{o/p-AFE}}{2N\cdot DR}
\end{aligned}
\end{equation}
$P_{laser}$ here represents the total electrical power consumed by the optical source (either a single comb laser source or multiple sources generating all the desired input wavelengths). To better study the energy efficiency based on a targeted signal resolution, $n_{i/p}$, the power consumption of the input laser source is formulated as a function of the optical power reaching the PDs at the output AFE as given in (\ref{laser_energy_efficiency_equation_mrr}).
\begin{equation} \label{laser_energy_efficiency_equation_mrr}
\begin{aligned}
  P_{\scalebox{.6}{laser}}&= \frac{10^{\frac{\eta_{\scalebox{.4}{WG[dB]}}N(d_{\scalebox{.4}{MRR}})}{10}}N}{\eta_{\scalebox{.4}{SMF}}\eta_{\scalebox{.4}{EC}}IL_{\scalebox{.4}{i/p-MRM}}(OBL_{\scalebox{.4}{MRM}})^{N-1}(EL_{\scalebox{.4}{splitter}})^{log_2N}} \\
  & \times 
  \frac{P_{\scalebox{.4}{PD-opt}}}{\eta_{\scalebox{.4}{WPE}}IL_{\scalebox{.4}{weight-MRR}}(OBL_{\scalebox{.4}{weight-MRR}})^{N-1}IL_{\scalebox{.4}{penalty}}}
\end{aligned}
\end{equation}
 where $d_{\scalebox{.4}{MRR}}$ represents the gap between the centers of two adjacent microrings and is dictated by the thermal crosstalk which should be taken for design considerations. A $d_{\scalebox{.4}{MRR}}$ of $15\mu m$ has been shown to be sufficient to avoid thermal crosstalk in a photonic switch implementation \cite{Ryotaro_Konoike_et_al_Ultra-compact_silicon_photonics_switch_with_high-density_thermo-optic_heaters}. We assume a $d_{\scalebox{.4}{MRR}}$ of $20~\mu m$ in this work for an optimistic realization of the system with $R_{MRR}=6~\mu m$. 

\begin{figure*}
\centering\includegraphics[width=0.709\linewidth]{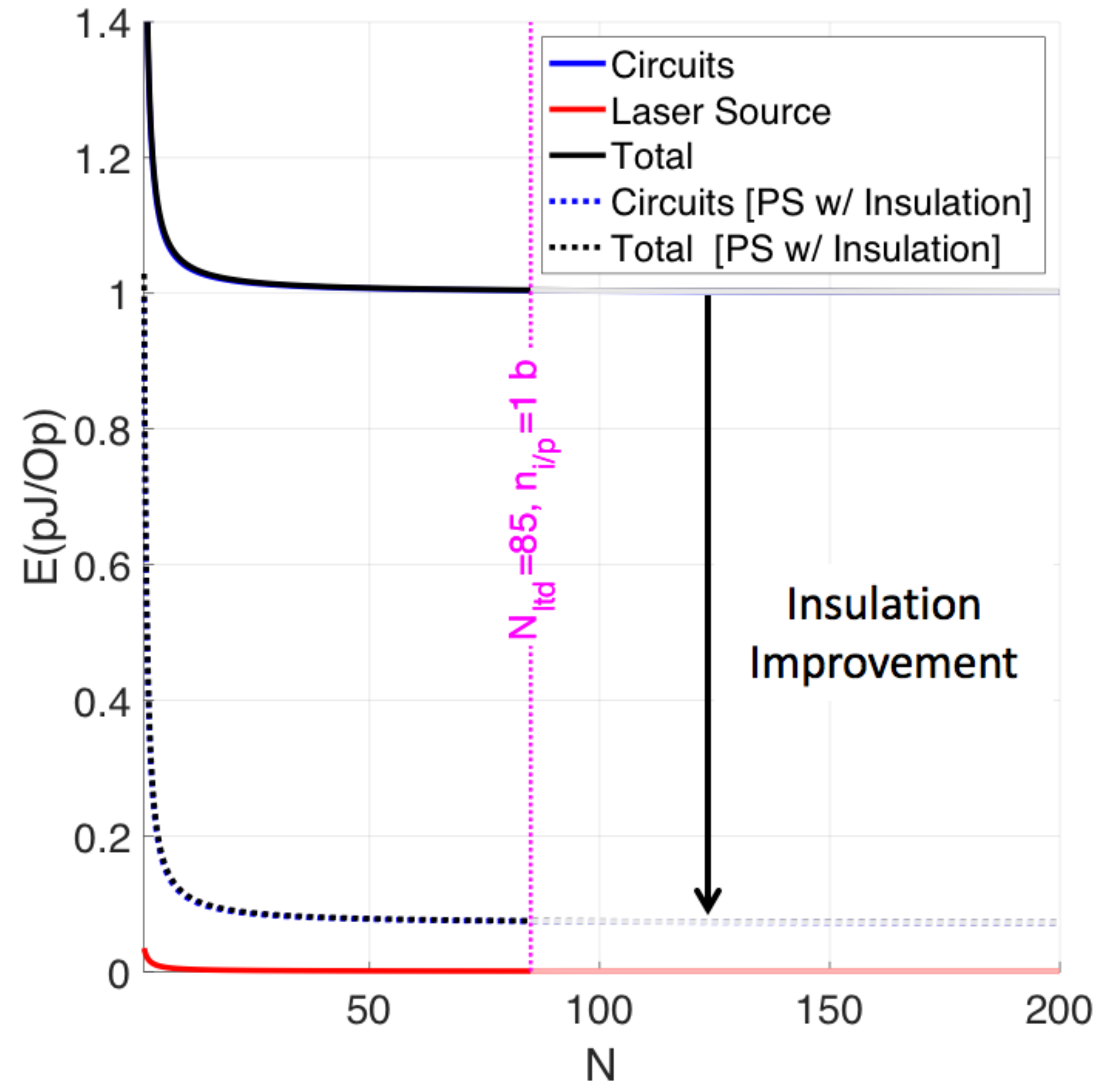}
\caption{Total energy efficiency (pJ/OP) for an $N\times N$ MRM implementation with PD responsivity $R=1.2~A/W$ and binary resolution, $n_{i/p}=1b$. Energy efficiency improves as the network scales up due to the increased number of operations. Improving the tuning efficiency with insulation has significant enhancement on the overall energy efficiency.}
\label{mrr_energy_efficiency_ILmrr_laser_electronic}
\end{figure*}

Since the number of operations scale quadratically with the weight vector size, $N$, while the energy consumption of the modulators and AFE increases linearly as can be seen in (\ref{mrr_based_energy_efficiency}), the overall energy efficiency improves with scaling as shown in Fig.~\ref{mrr_energy_efficiency_ILmrr_laser_electronic}. We assume similar MRM energy efficiency for multilevel signaling as PAM2, $\sim0.3pJ/b$,  \cite{Sajjad_moazeni_Vladimir_A_40Gbps_PAM4_Tx_based_on_a_ring_resonator_optical_DAC_in45nm_SOI_CMOS} which takes into account both the contribution of the modulator driver as well as the serializers. Therefore, the energy efficiency for $2b$, $3b$ and $4b$ input MRM drivers are estimated in our calculation as $\sim0.6pJ$, $0.9pJ$ and $1.2pJ$ per symbol, respectively.




MRMs offer smaller footprint and lower input capacitance which leads to significant reduction in their driving power. However, they are also more sensitive to fabrication mismatch and thermal drift which entails the need to use heaters for calibration across a wide spectral range. Excluding heater power, the power consumed by a closed loop controller implemented for a low-power WDM topology is $\sim 0.2~mW$ \cite{Xuezhe_Zheng_A_high_speed_tunable_silicon_photonic_ring_modulator_integrated_with_ultra_efficient_active_wavelength_control}. The average energy efficiency for state-of-the-art MRR heaters on an SOI platform is $\sim 20~mW/\pi$
\cite{Fuwan_Gan_Maximizing_the_Thermo_Optic_Tuning_Range_of_Silicon_Photonic_Structures, Hasitha_Jayatilleka_Photoconductive_heaters_enable_control_of_large_scale_silicon_photonic_ring_resonator_circuits, Adil_masood_Wim_Bogaerts_Comparison_of_heater_architectures_for_thermal_control_of_silicon_photonic_circuits}.
The scaling of the power consumed by the heaters with $O(N^2)$ degrades the overall energy efficiency of the network. Fig.~\ref{mrm_energy_efficiency_sensitivity_precision} shows the total energy efficiency and scaling limit for various  input  resolutions  considering  thermo-optic  phase  shifters with and without insulation. We assume power consumption values, $P_{heater}$, of $2.8~mW$\cite{Adil_masood_Wim_Bogaerts_Comparison_of_heater_architectures_for_thermal_control_of_silicon_photonic_circuits} and $40~mW$\cite{Hasitha_Jayatilleka_Photoconductive_heaters_enable_control_of_large_scale_silicon_photonic_ring_resonator_circuits} for phase shifters with and without insulation, respectively, to provide phase shift of one free spectral range (FSR). 

\begin{figure*}[h!]
\centering\includegraphics[width=0.910\linewidth]{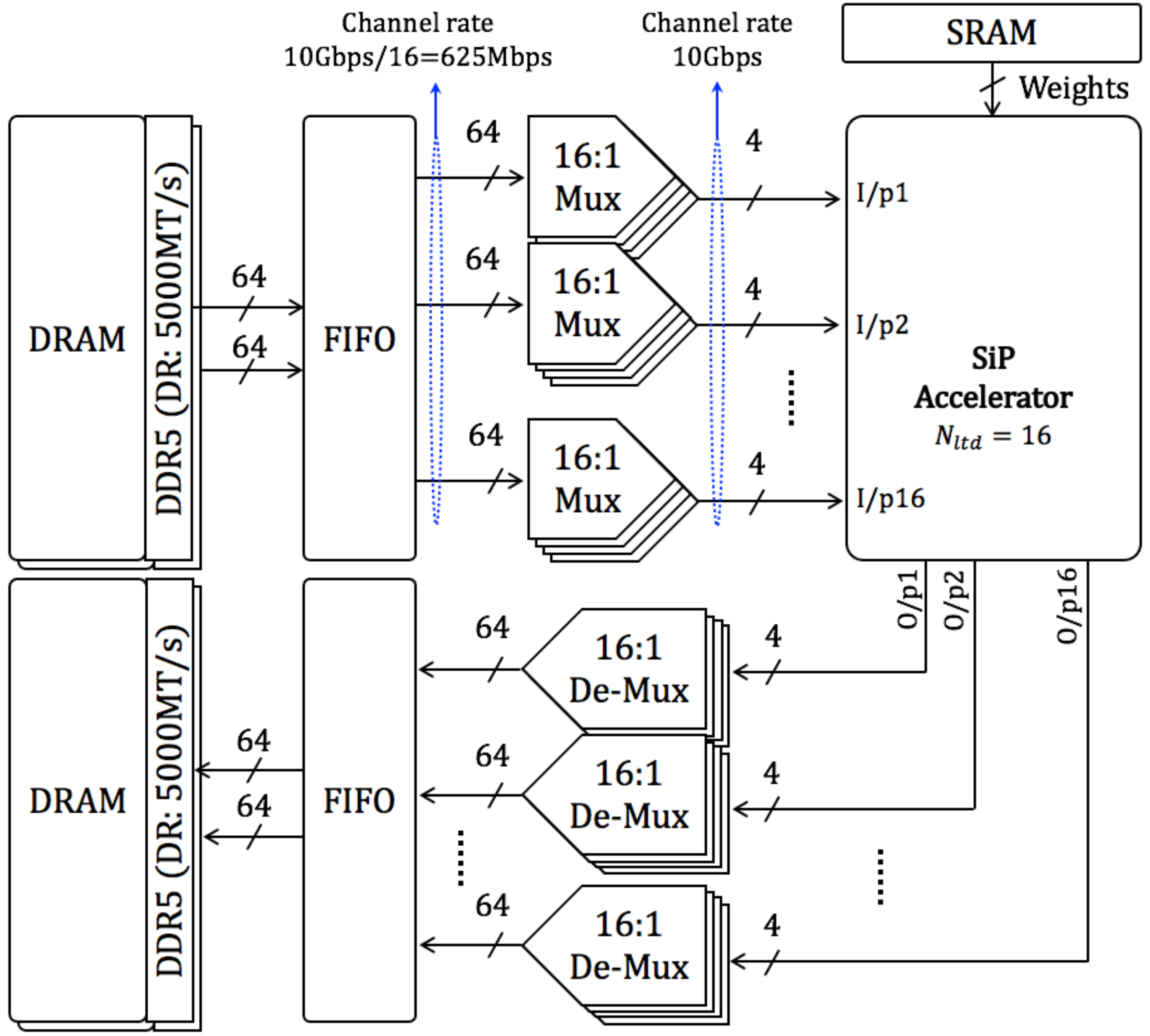}
\caption{Input data fetch for a SiP implementation with $n_{i/p}=4b$. FIFO and serializers are used to address the high-speed throughput requirement of SiP accelerators.}
\label{fig_data_fetch}
\end{figure*}

Driving inputs at high speeds entails the need to multiplex data fetched from the memory 
as shown in Fig.~\ref{fig_data_fetch}.  
As per the calculations in section \ref{section_MZM_energy_efficiency}, $P_{mem-interface}$ is taken as $5.77~mW$ for the energy efficiency calculation of input or output memory interfacing circuits.

\subsection{Scaling Limitations} \label{MRR_Tradeoffs}

\begin{figure*}
\centering\includegraphics[width=0.810\linewidth]{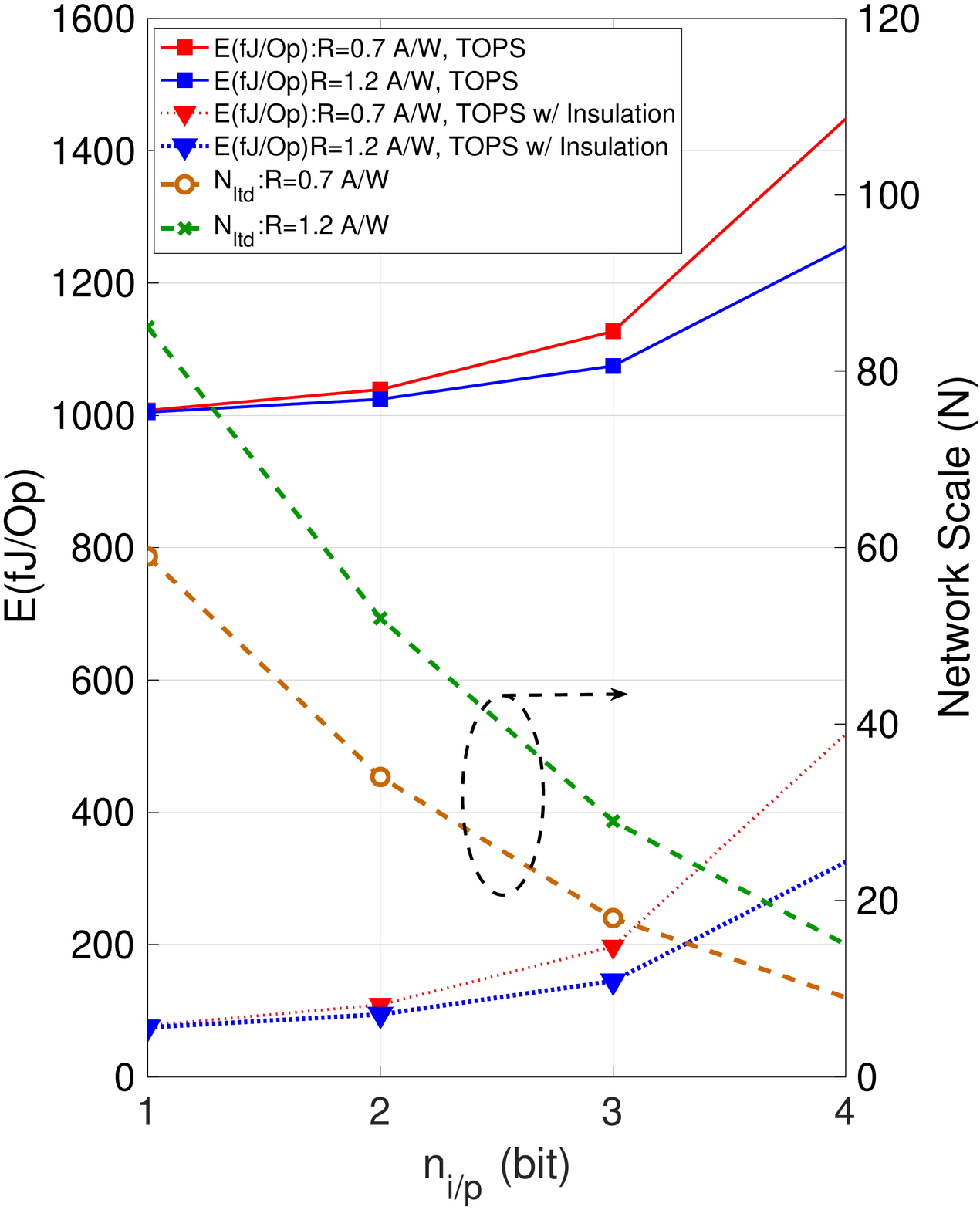}
\caption{Total energy efficiency and scaling limit of an MRM based network with an input resolution of $n_{i/p} = \{1b, 2b, 3b, 4b\}$ considering thermo-optic phase shifters with and without insulation at the scaling limit, $N_{ltd}$, where the laser rated power output (10~dBm) is reached. Implementing networks with higher resolution requires scaling down the network to improve the SNR at the AFE. This degrades the energy efficiency due to the reduced number of operations.}
\label{mrm_energy_efficiency_sensitivity_precision}
\end{figure*}

It can be inferred from Fig.~\ref{mrm_energy_efficiency_sensitivity_precision} that it is feasible to implement vector matrix multiplication using MRR based networks with sizes scaling up to $N = 85$. For $n_{i/p} = 2b$ or above, the network size is within the maximum number of microrings permitted for WDM implementations due to FSR limitations and crosstalk \cite{Alex_Tait_Microring_Weight_Banks}. Attempting to engineer the MRR’s dimensions and coupling ratio compromises the quality factor which degrades the channel spacings. This translates to a limit in the MRR vector size of  $N<FSR/\Delta \lambda$. Channel spacings are typically set according to the amount of acceptable crosstalk between channels (interchannel interference). As an example, for a $50~nm$ transmission window with channel spacings of $0.8~nm$, the maximum number of channels is $62$ \cite{Hasitha_Jayatilleka_Crosstalk_in_SOI_Microring_Resonator_Based_Filters, Alex_Tait_Microring_Weight_Banks}. Thus, for $n_{i/p} = 1b$, FSR may set a limitation to the overall network size.
  
Using series coupling to increase the filter order has been experimentally shown to reduce both interchannel and intrachannel crosstalks, thus, maximizing the filter finesse and the channel count \cite{Hasitha_Jayatilleka_Crosstalk_in_SOI_Microring_Resonator_Based_Filters}. However, this comes at the expense of higher footprint, lower drop port transmission and extra tuning power. To cascade several MRRs for MAC operations, it is necessary to maintain channel spacings to avoid the adjacent weight-dependent cross-talk. The number of channels that can be supported by optimized MRRs with finesse of 368 and 540 are calculated to be 108 and 148, respectively \cite{Alex_Tait_Microring_Weight_Banks,Qianfan_Xu_Silicon_microring_resonators_with_1.5_um_radius,Aleksandr_Biberman_Ultralow_loss_silicon_ring_resonators}. 

Two-point coupling scheme has been proposed to address the post-fabrication correction of MRM spectral features for large-scale MRM implementations \cite{Hossam_Shoman_Compact_wavelength_and_bandwidth_tunable_microring_modulator}. Although it mitigates the secondary resonances of an MRM and doubles the FSR, an extra micro-heater is introduced to correct for the coupling which increases the power consumption. Another attempt to achieve an FSR-free filter has been demonstrated using tunable couplers along with modified vernier filters that use higher-order coupled MRRs \cite{Maziyar_Milanizadeh_OSA2020_FSR_free_filter_with_hitless_tunability_across_C_L_telecom_band}. However, this topology is associated with penalty in terms of design complexity, increased footprint and tuning power. 

Introducing contra-directional coupling (CDC) in a microring combines the wavelength selectivity of the CDC with the compact feature size of the MRR, thus reaping the advantages of both and providing an FSR-free response \cite{Nourhan_Eid_Nicolas_Jaeger_FSR_free_MRR}. Implementing this design technique allows the potential use of several channels in MRR-based accelerators. This comes with a tradeoff of using extra heaters in the CDC and in the region of the MRR that does not include corrugated structures. 

As shown in Fig.~\ref{mrm_energy_efficiency_sensitivity_precision}, the optimum energy per operation of binary SiP networks based on MRRs ($\sim75~$fJ) is obtained at $N = 85$ for PD responsivity of $R = 1.2~$A/W. It can also be shown that reducing the power consumption of weight tuning circuits by one order of magnitude improves the energy efficiency by roughly one order of magnitude as well. Compared to their 8b digital CMOS counterpart, the energy efficiencies for SiP MRM MACs (using low-power thermo-optic phase shifters with insulation and 1.2~A/W PD responsivity) at $n_{i/p}$ = 1b to 4b resolutions are $~2.6\times$ to $~13\times$  worse. In comparison to MZM-based implementations, MRR-based implementations can have $1.8\times$ bigger network scale and achieve $1.3\times$ lower energy consumption per operation. Similar to MZM-based implementations, MRR MACs are performed at a $2N\alpha f_{CLK-OPT}/f_{CLK-CMOS}\times$ higher throughput than its digital CMOS counterparts. 

\begin{figure*}
\centering\includegraphics[width=1.0595\linewidth]{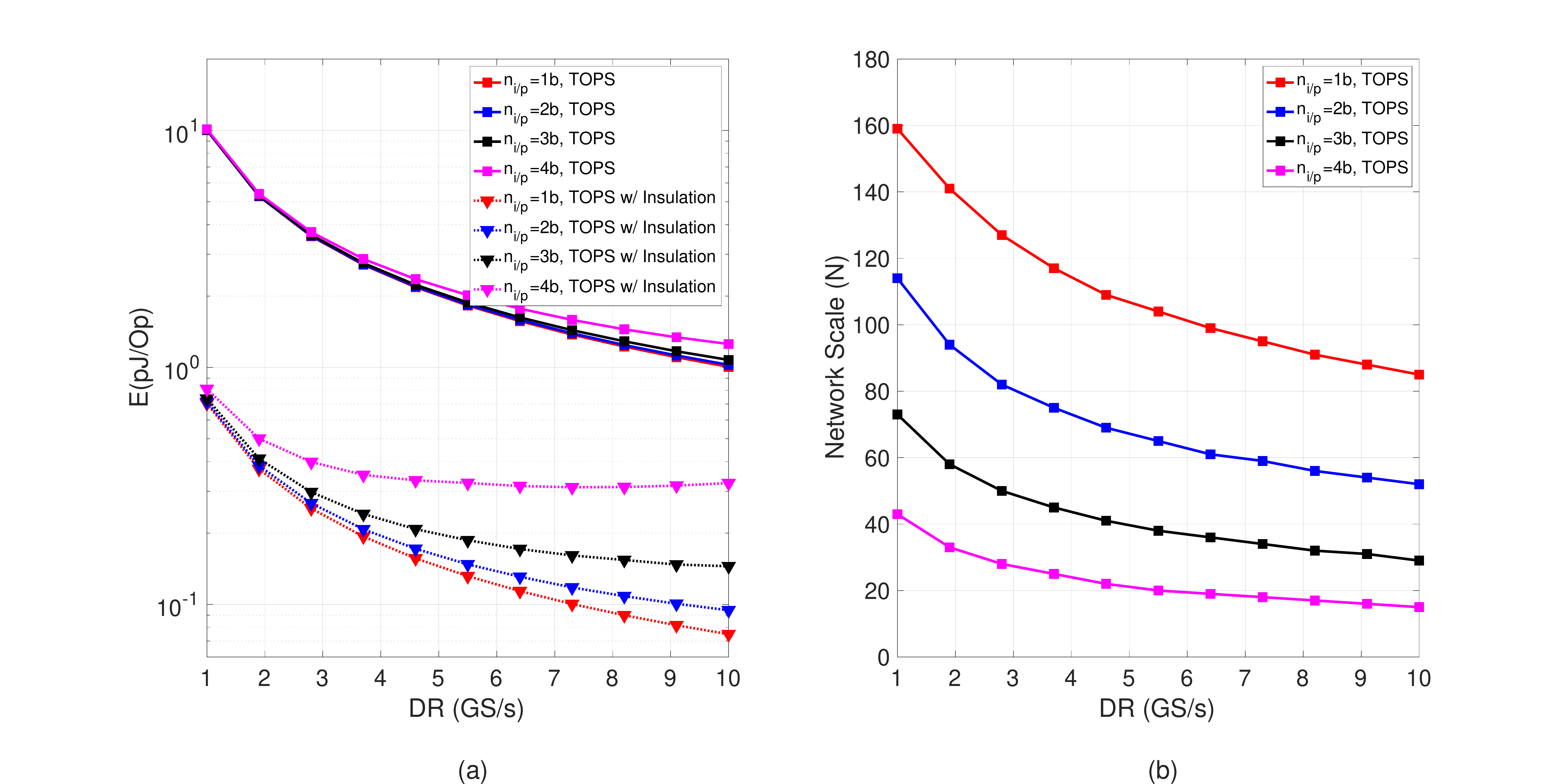}
\caption{(a) Total energy efficiency and (b) MRM Network sizes for  $R=1.2~A/W$, $DR=1-10~GS/s$. Operating at higher data rates helps reduce the contribution of the static power consumption of thermo-optic phase shifters to the energy efficiency. Network scales up at lower data rates due to the SNR improvement.}
\label{fig_E_N_vs_DR_mrr}
\end{figure*}

Fig.~\ref{fig_E_N_vs_DR_mrr} shows the energy efficiency and scaling at lower data rates. Similar to MZM implementations, reducing the data rate degrades the energy efficiency while scaling up the network size due to the reduced noise levels at the AFE.

\section{Research Opportunities} 

\begin{figure*}
\centering\includegraphics[width=0.810\linewidth]{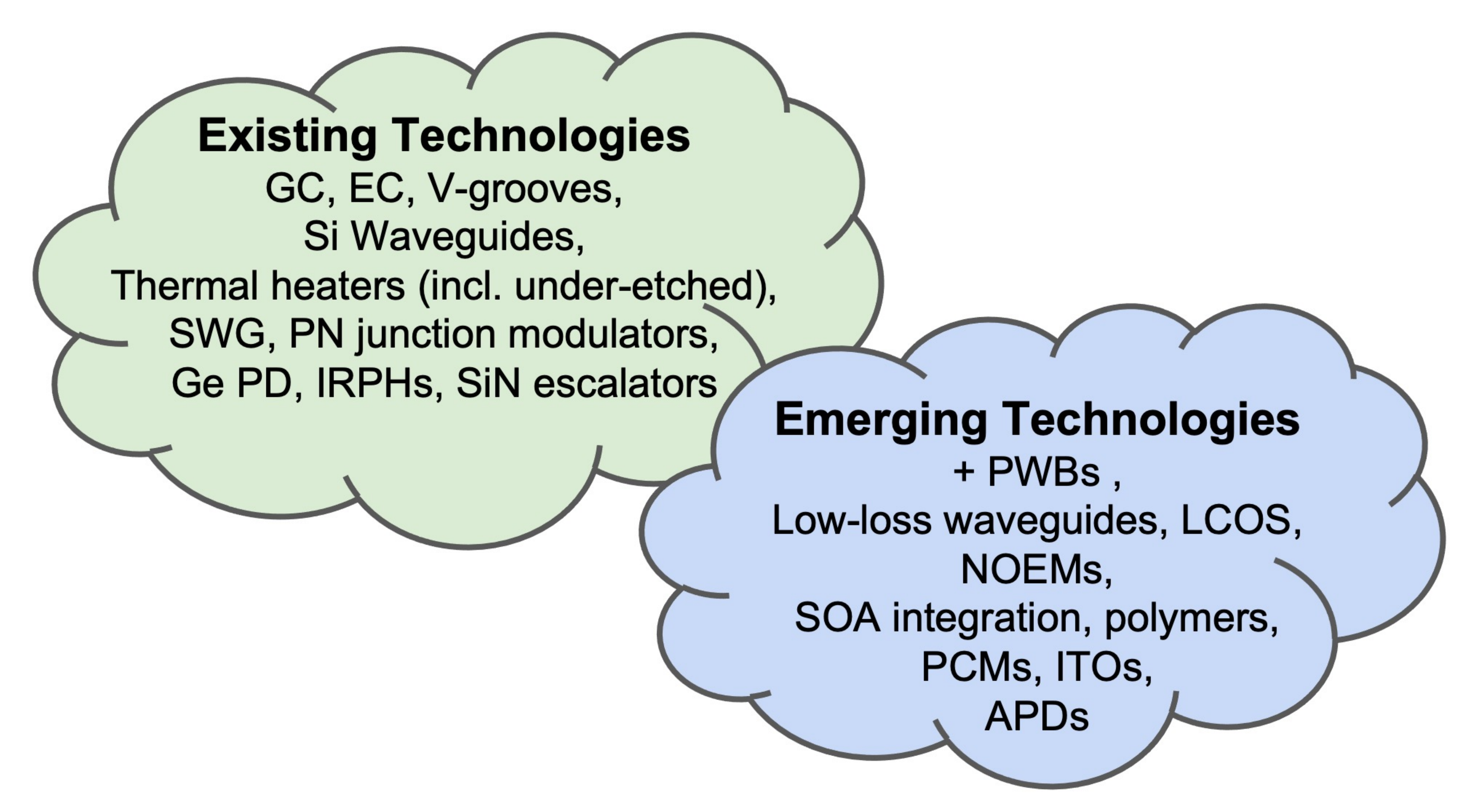}
\caption{\label{fig_SiP_versions} Evolution of SiP technology from current (1.0) to the next generation (2.0). Evolution of the existing SiP technologies - which comprise of grating couplers (GCs), Edge Couplers (ECs), V-grooves, silicon waveguides, thermal heaters, sub-wavelength gratings (SWGs), PN junction modulators, germanium photodetectors (Ge PDs), in-resonator photoconductive heaters (IRPHs), silicon nitride (SiN) escalators - to include emerging technologies such as photonic wire bonds (PWBs), low-loss waveguides, liquid crystal on silicon (LCOS), nano-opto-electromechanical systems (NOEMs), semiconductor optical amplifiers (SOAs), polymers, phase change materials (PCMs), indium tin oxide (ITOs) and avalanche photodetectors (APDs).}
\end{figure*}

As summarized in sections \ref{section_MZM_based_SiP_Implementation} and \ref{Section_MRR_based_SiP_Implementation}, SiP accelerators operate at much higher speed and lower latency than their CMOS counterparts. Nevertheless, it is further desired to improve the size of the MAC networks in SiP, especially for neural network applications, and improve the energy efficiency. 
There has been several promising research in the field of SiP. Classifying the existing commercial SiP technology as the first generation, we describe several emerging technologies that will make up the next generation of SiP Fig.~\ref{fig_SiP_versions} summarizes the advancements in SiP that can be leveraged by SiP-based accelerators to reduce optical loss, improve the energy efficiency and incorporate heterogeneous integration techniques for performance improvement. 

\subsection{Optical Loss Reduction}
As described in sections \ref{section_MZM_based_SiP_Implementation} and \ref{Section_MRR_based_SiP_Implementation}, optical losses limit the scalability of the SiP technology. Losses must be minimized at the coupling interfaces and in the components. PWB is one way to ensure efficient coupling between the chip and the optical fiber with insertion loss $\sim 1~$dB with negligible variation \cite{vanguard_photonic_wirebonding}. Passive alignment to SMF optical fibers can be accomplished using V-grooves arrays. Such fiber to chip self-alignment has been shown to have coupling efficiency of $\sim -1.3$ dB \cite{Tymon_Barwicz_IBM_A_Novel_Approach_to_Photonic_Packaging_Leveraging_Existing_High-Throughput_Microelectronic_Facilities}. In another demonstration, coupling losses as low as $\sim 0.5$ dB and $\sim 0.35$ dB have also been reported for passive and active alignments, respectively \cite{Saeed_Fathololoumi_1.6Tbps_Silicon_Photonics_Integrated_Circuit_for_Co-Packaged_Optical-IO_Switch_Applications}.

For scaling up the networks, the optical signal attenuation can be compensated by using SOAs. On-chip SOAs can be utilized to pre-amplify the input signal and also exploited as weight matrix elements to provide weights magnitudes $>1$ \cite{Bin_Shi_Deep_Neural_Network_Through_an_InP_SOA_Based_Photonic_Integrated_Cross_Connect}. However, the non-linear gain-current curve entails a need for calibration. 

Improving the responsivity of the AFE is yet another way to tolerate the optical losses. It relaxes the need to increase the laser power to compensate for the losses. The high multiplication gain and responsivity of APDs have been shown to improve the sensitivities of optoelectronic receivers front-end \cite{Spoorthi_Nayak_10Gbps_18.8dBm_Fully_Integrated_Optoelectronic_Receiver_With_Avalanche_Photodetector}. Improving dark current and quantum efficiency by careful design of the APD geometry has been projected to improve the sensitivity of Si-Ge APD receivers up to $-29~$dBm at 12.5~Gb/s \cite{Zhihong_Huang_Si_Ge_APD} as compared to $-18.5~$dBm for Ge PIN detectors\cite{Jiho2010_Ge_on_Si_PD}. Limiting the bandwidth of the AFE and using equalization techniques \cite{ShekharCICC13} such as continuous time linear equalization (CTLE) and decision feedback equalization (DFE) \cite{IBM_3.3_monolithic_PD_CMOS_preamplifier} can reduce the input referred noise of the AFE and further improve the sensitivity.

\subsection{Improving energy efficiency}

Commercial CW lasers suffer from low WPE in the range of $\sim 1\%-10\%$, which impacts the energy efficiency on the system \cite{A._J._Zilkie_Power-efficient_III-V/Silicon_external_cavity_DBR_lasers, Shinsuke_Tanaka_et_al_High-output-power_single-wavelength_silicon_hybrid_laser_using_precise_flip-chip_bonding_technology}. 
Hybrid-integrated silicon photonic lasers have been shown to provide $\sim 12.2\%$ WPE
\cite{Jin_Hyoung_Lee_12.2WPE_hybrid_laser}.

Although introducing on-chip CW lasers mitigates the coupling losses, the feasibility of using them, especially for networks using WDM, require wavelength stabilization and reflection cancellation \cite{Christopher_R._Doerr_Optical_isolator_using_two_tandem_phase_modulators, Hossam_JLT2021}.

Reducing the power consumption of the phase shifters in the weight matrix is a critical requirement given that their overall energy consumption scales quadratically with network size (Eq.~(\ref{mzm_based_energy_efficiency}) and Eq.~(\ref{mrr_based_energy_efficiency})). 
Thermo-optic phase shifters dissipate high power consumption given their resistive nature. Introducing trenches, undercuts and back-side substrate removal has been shown to improve the tuning efficiency of the rings by an order of magnitude with measured reported power consumption of $\sim 4~mW$ per FSR \cite{Po_Dong_Thermally_tunable_silicon_racetrack_resonators_with_ultralow_tuning_power, Xuezhe_Zheng_silicon_photonic_WDM_transmitter_using_off_chip_laser_sources,Adil_masood_Wim_Bogaerts_Comparison_of_heater_architectures_for_thermal_control_of_silicon_photonic_circuits}. However, thermal isolation and substrate removal exacerbates self-heating and must be taken into consideration while designing a CMOS controller \cite{chen_sun_vladimir_A_45nm_CMOS_SOI_Monolithic_Photonics}. 

Several post-fabrication schemes have been investigated to correct for the fabrication-induced variations. Reducing the process variations was investigated by patterning SiN on top of the Si waveguide to introduce field perturbations which effectively adjusts the optical path length \cite{Amir_Atabaki_sin_post_fabrication_trimming}. Another demonstrated technique relies on trimming using Ge ion implantation followed by laser annealing to tune MRR resonant wavelength across the whole FSR without introducing any excess loss. Its accuracy, CMOS-compatibility, and feasibility for wafer-scale correction renders it a potential technique to be utilized in optical neuromorphic implementations to reduce the tuning power \cite{Xia_Chen_Graham_reed_Germanium_implanted_photonic_devices_for_post_fabrication_trimming_and_programmable_circuits}.  

Alternatives such as nano-opto-electro-mechanical systems (NOEMS) \cite{Carl_Ramey_Lightmatter_Hotchips_Silicon_Photonics_for_Artificial_Intelligence_Acceleration, Yu_Feng_et_al_Performance_analysis_of_a_silicon_NOEMS_device_applied_as_an_optical_modulator_based_on_a_slot_waveguide} and liquid crystal on silicon (LCOS) \cite{Yufei_Xing_Wim_Bogaerts_Digitally_Controlled_Phase_Shifter_Using_an_SOI_Slot_Waveguide_With_Liquid_Crystal_Infiltration} have the potential to reduce the tuning power overhead significantly. The dynamic energy consumption of NOEMS was reported in the range of $0.13$~fJ and $0.32$~fJ for digital pulse signals \cite{Yu_Feng_et_al_Performance_analysis_of_a_silicon_NOEMS_device_applied_as_an_optical_modulator_based_on_a_slot_waveguide}, and is assumed as $\sim1$~fJ for our study. On the other hand, LCOS have been shown to dissipate power as low as $2$~nW \cite{Yufei_Xing_Wim_Bogaerts_Digitally_Controlled_Phase_Shifter_Using_an_SOI_Slot_Waveguide_With_Liquid_Crystal_Infiltration}.

\begin{figure*}
\centering\includegraphics[width=0.9\linewidth]{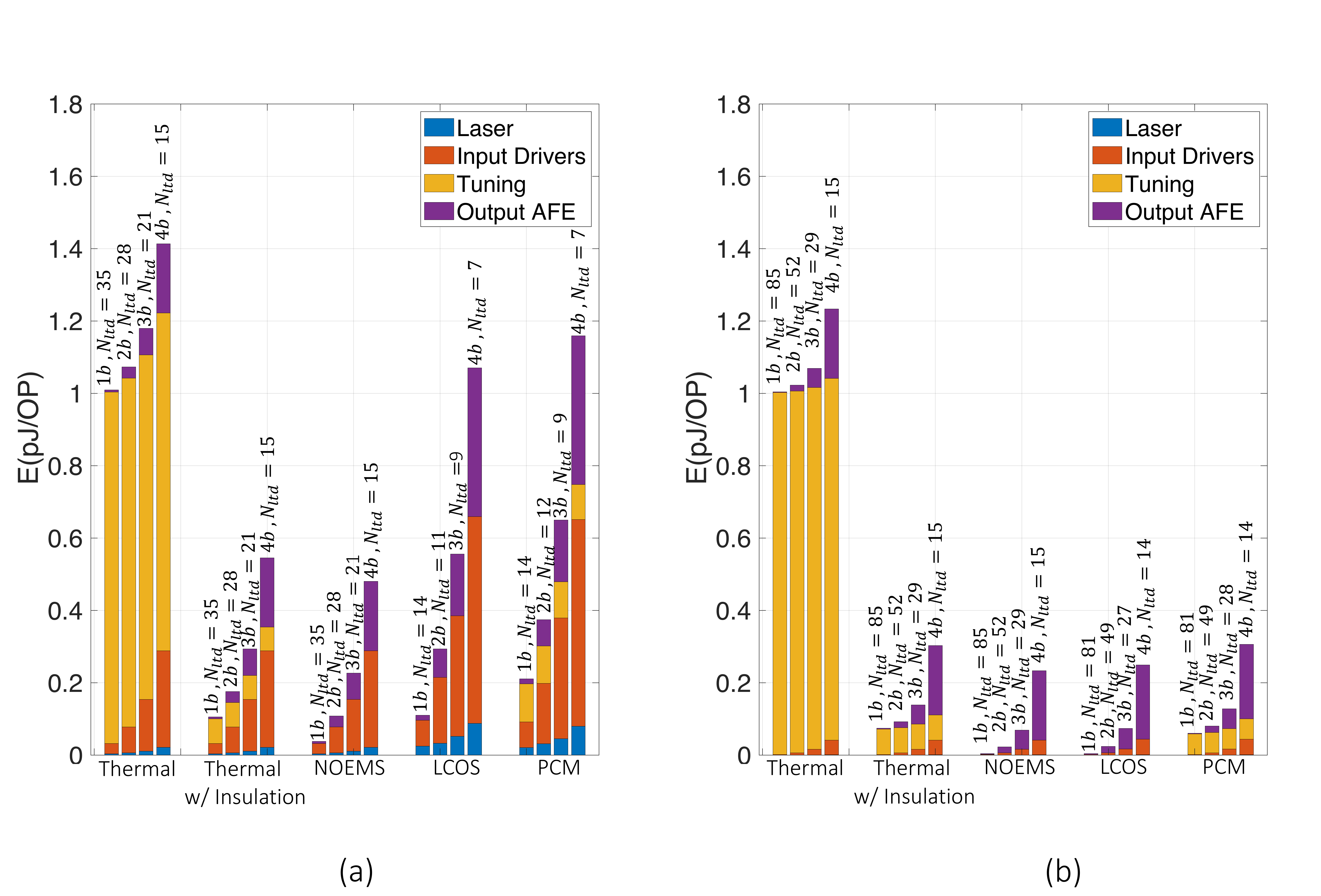}
\caption{Energy efficiency breakdown considering various weight tuning options for bit resolutions, $n_{i/p}=\{1, 2, 3, 4\}b$ for (a) an MZM-based implementation, and (b) an MRM-based implementation. Phase shifters with low insertion loss and static power consumption (e.g. TOPS with insulation and NOEMS) are good candidates to enhance the energy efficiency of SiP implementations. The high insertion loss of LCOS and PCM poses limitations on the network sizes for a target resolution. In addition, PCMs, with their large dynamic power consumption, are promising for MRM implementations provided a high weight reuse is possible.}
\label{mzm_mrr_prospects}
\end{figure*}

Phase change materials (PCMs) such as $Ge_2Sb_2Se_4Te (GSST)$ have been demonstrated as compact phase shifters in which the optical phase shift is obtained by tuning the state of the material from amorphous and crystalline \cite{Zhang2018, Nikhil2019}.
Being able to sustain their crystallization state with the absence of power renders them as good candidates for tuning low-speed weights in SiP implementations with no static power consumption. Given their compact sizes and non-volatile nature, the efficiency of implementing them in large-scale SiP networks is investigated as shown in Fig.~\ref{mzm_mrr_prospects}. Although not as lossy as PN phase shifters, PCMs have an $IL = 0.32~dB$ which is relatively high for cascaded phase shifters in a large-scale implementation\cite{Zhang2018}. This limits the network sizes for computation with several bits of resolutions. The resolution of the weights can be set by adjusting the level of crystallization of a PCM cell \cite{Carlos2015}.

The pulse energy consumption for writing and erasing levels 1-7 were reported in the range of $372pJ - 601pJ$ and $562pJ - 373 pJ$ \cite{Carlos2015}, respectively. Assuming the weights to be uniformly distributed, the average energy consumption, $E_{PCM}$, for setting the PCM to various weights can be calculated as in Eq.~(\ref{eq_E_PCM}), where $E_{A}$ and $E_{C}$ represent the pulse energy required to write (amorphization) and erase (crystallization) the first level ($L_1$), respectively. For levels, $L_i$, where $i>1$, $\Delta E_{A}$ and $\Delta E_{C}$ represent the average amount of energy required to transition to one level higher or a lower, respectively, with all levels assumed to be equally spaced for the sake of simplicity. 
\begin{equation}
\begin{aligned}
    E_{PCM}&=\frac{2^{n}-1}{2^{2n}}(E_{A}+E_{C}) \\
    &+\frac{(1/3)(2^{2n}-1)2^{n-1}-(2^{n}-1)}{2^{2n}}(\Delta E_{A}+\Delta E_{C})
\end{aligned}
\label{eq_E_PCM}
\end{equation}
Assuming $E_{A}=372~$pJ, $E_{C}=373~$pJ, $\Delta E_{A}=(601-372)/(2^{n}-2)~$pJ and $\Delta E_{C}=(562-373)/(2^{n}-2)~$pJ, the estimated average energy consumption for a PCM phase shifter with $n=\{1, 2, 3, 4\}b$ equals $\{186, 231, 165, 121\}~$pJ. For phase shifters with zero static power dissipation, the dynamic energy consumption is divided by the number of times weights have been reused for vector matrix multiplication. This is done by introducing a weight reuse factor, $\alpha_{w}$, such that $P_{mat-tuning}=P_{NOEMS,PCM}/\alpha_w$, where $\alpha_w$ typically ranges between $2^{6}$ to $2^{18}$ in general matrix multiplications (GEMMs) \cite{Nvidia_Docs}. 
A value of $\alpha_{w}=4096$ is chosen for the calculation of energy efficiencies in this work. For networks where the weight reuse is low, the contribution of the dynamic energy per operation can be considerably higher for PCM than all the other weight tuning alternatives, degrading the energy efficiency by orders of magnitude.

Fig.~\ref{mzm_mrr_prospects} shows the energy efficiency breakdown for both the MZM-based and MRR-based architectures for weight tuning that rely on thermo-optic phase shifters without and with insulation \cite{Adil_masood_Wim_Bogaerts_Comparison_of_heater_architectures_for_thermal_control_of_silicon_photonic_circuits}, NOEMS, LCOS and PCM. For each implementation, energy calculations are based on the network scales that can satisfy the SNR requirements to compute with bit resolutions, $n_{i/p}=\{1, 2, 3, 4\}b$. The insertion loss for a $35~\mu m$ long LCOS used to realize a $\pi$ phase shift is taken as 0.35~dB \cite{Yufei_Xing_Wim_Bogaerts_Digitally_Controlled_Phase_Shifter_Using_an_SOI_Slot_Waveguide_With_Liquid_Crystal_Infiltration}. For MZM implementations, thermo-optic phase shifters with insulation seem currently attractive for energy efficiency and network size. For a large weight reuse factor, NOEMS-based phase shifters promise further energy reduction. For MRM implementations, similar conclusions can be drawn except that LCOS-based phase shifters also seem promising.


For both MZM-based and MRR-based architectures, it is evident that opting for matrix weight tuning alternatives with almost zero power consumption significantly improves the total energy efficiency, with values approaching $< 100$~fJ/Op for both architectures. Further research is still needed to demonstrate the feasibility of these approaches in high volume production to realize such energy efficiency regime.  

\begin{figure*}
\begin{equation}
\resizebox{.9\hsize}{!}{$
SNR = \frac{(RGP_{o/p})^2}{\left[\sqrt{\left(2q(RGP_{o/p}+I_d)+\frac{4kT}{R_L}+2\rho_{ASE}R^2GP_{o/p}+\rho_{ASE}^2R^2(2B_o-B_e)+R^2P_{o/p}^2RIN\right)}+\sqrt{(2qI_d+\frac{4kT}{R_L}+\rho_{ASE}^2R^2(2B_o-B_e))}\right]^2B_e}
$}
\label{SNR_SOA}
\end{equation}
\end{figure*}

\subsection{SOA Cascadability}

SOAs are used to amplify optical signals over a given spectrum and can be implemented off-chip or using hybrid integration. Cascading SOAs has been conventionally used to restore signal levels in interconnect links and has been recently explored in deep neural network implementations \cite{Bin_Shi_Deep_Neural_Network_Through_an_InP_SOA_Based_Photonic_Integrated_Cross_Connect}.

Although SOAs help compensate for the optical losses to increase the scale of the network, they contribute significantly to the energy consumption. In addition, they suffer from several downsides, including the noises produced due to the optical amplification, the ripples in their gain spectrum, and amplification nonlinearity \cite{connelly_2011, sasikala_chitra_2018, BANEY_2000}. The major noise component is attributed to the amplified spontaneous emission (ASE) of photons towards the input and output of an SOA \cite{laser_diodes_book, BANEY_2000}. Therefore, the build-up of ASE noise due to the cascade degrades the SNR of the optical signal reaching the output \cite{connelly_2011, BANEY_2000}. 

To investigate the effect of incorporating SOAs on the scalability of the network as well as the received signal resolution, the overall SNR is quantified in Eq.~(\ref{SNR_SOA}), with an SOA gain value as $G = 17~$dB\cite{Rongqing_Hui_2020, zhiwei_photonics}. $B_o$ and $B_e$ stand for the optical bandwidth of the amplifier and electrical bandwidth of the AFE, respectively. $\rho_{\scalebox{.5}{ASE}}$ represents the ASE noise and is calculated as:
\begin{equation}
    \rho_{\scalebox{.5}{ASE}} = 2N_{\scalebox{.5}{SOA}}n_{sp}\frac{hc}{\lambda}(G-1)
\end{equation}
where $N_{\scalebox{.5}{SOA}}$ and $n_{sp}$ represent the number of SOAs used in the network and the spontaneous emission factor of the optical amplifier, respectively.

\begin{figure*}[h]
\centering\includegraphics[width=0.710\linewidth]{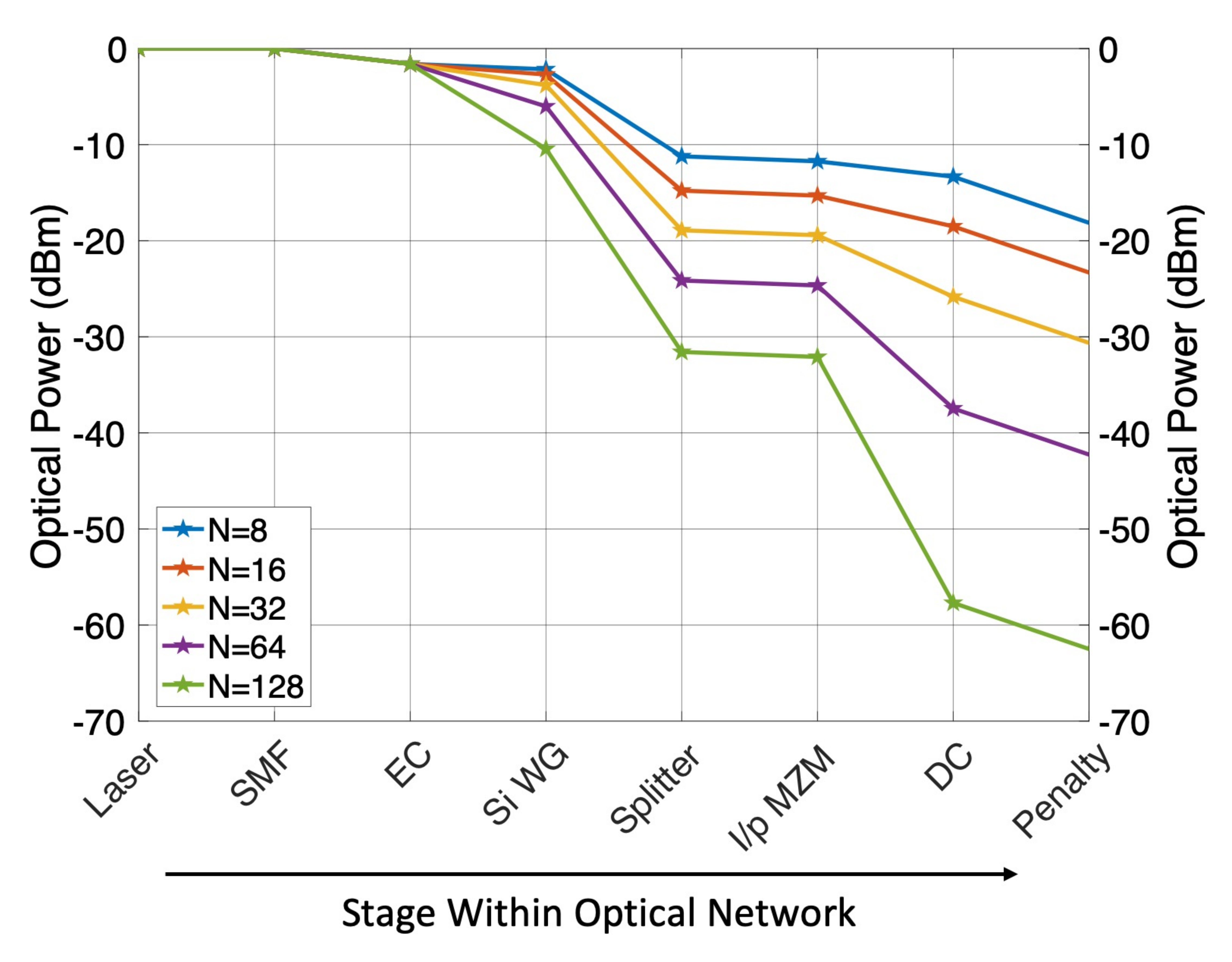}
\caption{Optical link budget analysis for MZM-based matrix sizes of N=8, 16, 32, 64, 128 for $R = 1.2~A/W$ and $DR = 10~GS/s$.}
\label{optical_link_budget_N}
\end{figure*}

\begin{figure*}
\centering
\includegraphics[width=0.710\linewidth]{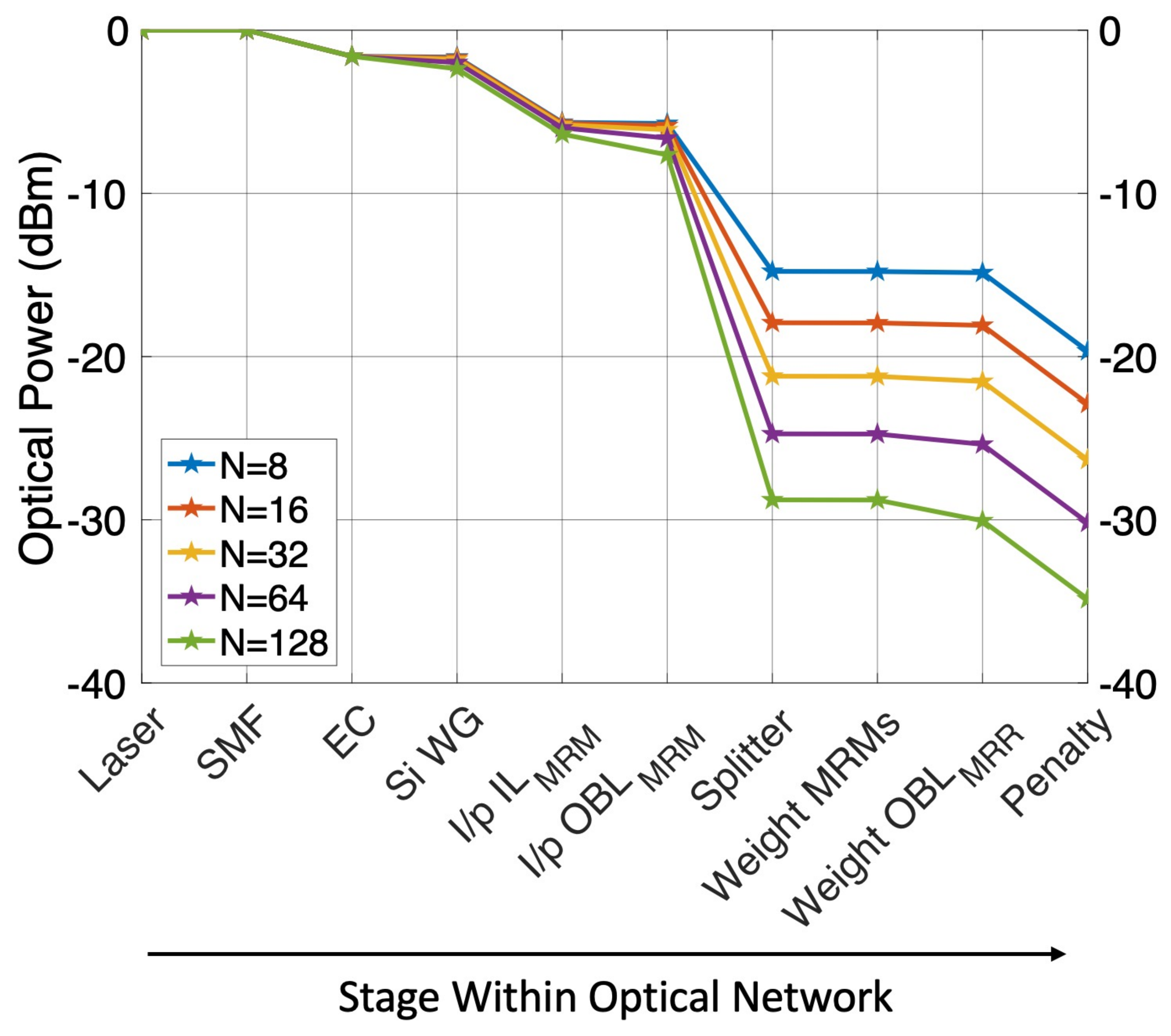}
\caption{Optical link budget analysis for MRR-based matrix sizes, N=8, 16, 32, 64, 128 for R=$1.2~A/W$ and DR=$10~GS/s$.}
\label{optical_link_budget_N_mrr}
\end{figure*}

The parameters given in Table~\ref{table_SNR_calculations} are used for calculating the SNR and $N_{\scalebox{.5}{SOA}}$. For a target network resolution, $n_{i/p}$, the number of SOAs are calculated based on the resultant SNR whose signal and noise power values vary with the network scale. The SNR is calculated as in Eq.~(\ref{eq_snr}).
\begin{equation}
    n_{\scalebox{.8}{i/p}} = \frac{SNR[dB] - 1.76}{6.02}
    \label{eq_snr}
\end{equation}
As can be inferred form Fig.~\ref{optical_link_budget_N} and Fig.~\ref{optical_link_budget_N_mrr} for an SOA-less network, the SNR degrades since the signal optical intensity at the AFE $P_{o/p}$ is attenuated with the scale of $N$. Incorporating an SOA helps replenish the signal intensity. SOAs can be added in the network before the SNR degrades below the threshold for a given resolution due to the insertion losses. However, the signal dependent noises are amplified as well which do not align in favor of the network resolution. For calculating the resolution, the AFE is assumed to tolerate signals with maximum $P_{opt}$ intensity as high as $10~dBm$, beyond which the number of SOAs are limited. Regions with no SOAs at the right side of Fig.~\ref{fig_MZM_N_vs._n_vs._Nsoa} and Fig.~\ref{fig_MRM_N_vs._n_vs._Nsoa}, shown in appendix A, represent regions where the desired SNR cannot be achieved, indicating an infeasible resolution for a given scale. 



SOAs require introducing III-V or II-VI compound semiconductor materials to the SiP platform which is non-compliant with the standard SiP CMOS foundry runs. Back propagating ASE noise from the SOA emphasizes the need to employ an optical isolator for the input laser source \cite{connelly_2011}. Use of narrowband filters are also possible to reduce the out-of-band noise \cite{Bin_Shi_Deep_Neural_Network_Through_an_InP_SOA_Based_Photonic_Integrated_Cross_Connect, connelly_2011}, but these further reduce the maximum channel count, thus limiting the scaling of the network. To be used for linear analog computations, SOAs should deliver gains that are independent of the input intensities. Cross-gain modulation (XGM) is one type of non-linearity observed in SOA amplifiers in which the combination of all the input intensities impacts the gain of a single channel \cite{Chauhan2018CrossGM, Shuangmei2004}. 

The aforementioned SOA limitations should be addressed in order to maintain the linearity of the weight matrix and allow further scaling of networks. Designing highly efficient SOAs has the potential to increase the size of the networks. However, improving the network energy efficiency requires that the SOAs have low injection current and high optical signal to noise ratio. Recent attempts to use SOAs for optical networks have reported power consumption of 42~mW per SOA which leads to an energy consumption of $\sim$ 4.2~pJ/Op at a $DR=10~$GS/s \cite{Bin_Shi_Deep_Neural_Network_Through_an_InP_SOA_Based_Photonic_Integrated_Cross_Connect}. To carry out 4 weighted additions, 16 SOAs were used for the weight tuning, along with extra SOAs for optical pre-amplification and input selection. A crosstalk of 0.6~dB was reported for the SOAs even for a small-scale circuit implementation of arrayed waveguide grating (AWG) filter, which necessitated the use of feedback loops for gain calibration. For accelerators with SOA integration, the contribution to the overall network energy efficiency scales with $O(N^2)$, setting the efficiency to the pJ/Op regime.

\section{Conclusion}
We describe the behavior of MZM and MRM based SiP implementations for MAC accelerators based on today's SiP 1.0 technology. Both MZM and MRM implementations share similar optical and electrical challenges. In comparison to digital CMOS accelerators, SiP implementations have relatively higher energy consumption and operate at lower bit resolutions. In addition, they cannot be scaled to large network sizes because of the optical losses. Implementing MACs using SiP has two distinct advantages\cite{Bhavin_2021}:
\begin{enumerate}
    \item Optical MAC operations can be scaled to frequencies at tens of GHz, whereas MACs in digital CMOS are limited to a few hundreds of MHz or at most GHz operating speeds. For tasks where memory access is not the bottleneck, such as inference with fixed weights, an optical implementation can reduce latency and improve the energy efficiency at such high speeds. Digital CMOS counterparts, on the hand, are limited by the clock frequency.
    \item Multiplication operations can be intrinsically implemented in parallel in which the analog nature of the computation allows all matrix operations to take place at the same time for each input fetch \cite{Tamir_parallel_vmm}. Therefore, optical MAC implementations can increase their throughput and improve the energy efficiency at such high speeds. MACs implemented with digital circuits in CMOS are limited by the wiring interconnect density.
\end{enumerate}

There are also some challenges of implementing MACs using SiP:
\begin{enumerate}
    \item The losses in optical circuits severely limit the size of the MAC networks that can be physically realized, in comparison to a digital CMOS implementation where signal gain and regeneration is easily available. This, in turn, limits the applications of SiP MACs.
    \item Although the power consumed in the optical MAC is small, when accounting for the losses and the power consumed by the laser and CMOS electronic circuits that drive and control the optical circuits, the energy efficiency is degraded.
    \item Unlike digital CMOS implementations which can support 16b/32b resolution, analog photonic MACs have a maximum demonstrated resolution of 8.5b \cite{zhang2021microring}. Nevertheless, such a resolution has been shown to be adequate for many inference tasks \cite{Itay_Hubara_Quantized_Neural_Networks,Edward_H._Lee_LogNet, Steven_K._Esser_Convolutional_networks_TrueNorth}.
    \item Achieving a high throughput optical network entails accessing data at high speed. High-speed input/output (I/O) data can be streamed in/out from/to an off-chip DRAM; the corresponding energy consumption in moving the data must be considered for the overall implementation of an accelerator\cite{Chris_Cole_2021}. The energy consumption in data fetch from off-chip DRAM is significantly large. However, for a given dataset, the DRAM associated penalty is similar for both optical and digital CMOS implementations. We exclude that penalty in our work. For weight-stationary implementations where weights do not need to change frequently, an on-chip SRAM can be used, which can be adequately large due to the limited network size of the photonic accelerators.
    \item Most of the low-loss phase shifters have a reconfiguration speed in the range of $\mu$s-to-ms, making the weight reconfiguration in photonic MACs significantly slower than their electronic counterparts. This limits the use of photonic MACs to weight-stationary systolic array implementations, where the incoming data is high-speed, but the weight does not get updated quickly. For other scenarios, a fine weight retuning can be done with high-speed plasma-dispersion phase shifters, where the loss is controlled due to the need for a fine weight tuning range only.
    \item To carry out optoelectronic computing with several bits of resolutions at high speed, CMOS or biCMOS drivers and transimpedance amplifiers (TIAs) are needed that must operate with multi-level pulse-amplitude modulation (PAM) signaling. Although many PAM2 ($1b$) and PAM4 ($2b$) transceivers have been demonstrated\cite{xiaotie_Wu_pam_si_ph_modulator, 56Gbps_PAM4_Tanaka2018}, higher levels of modulation require linear drivers and TIAs which are challenging to design at high speed and good energy efficiency\cite{AR_JSSC_A_Dual_Polarization_Silicon_Photonic_Coherent_Transmitter_Supporting_552_Gbs_wavelength}. However, if data rates in optical computing are limited to a few tens of GBaud, this challenge is surmountable.
    
    \item Packaging considerations in optics (e.g., laser, fiber and SOA attach) are far more challenging than the packaging considerations for electronic dies due to alignment accuracy and thermal management requirements \cite{Lee_2016_Photonic_packaging}.
\end{enumerate}

 The mesh-like interconnections of MZM-based implementations which extend the optical dynamic range at the AFE place a tradeoff between the output resolution and the power requirement of the laser. MZM drivers also consume higher electrical driver power because MZMs cannot be made very long due to the losses associated with their larger footprints. Therefore, the power consumption of the laser and high speed drivers are amortized with a limited network scalability. On the other hand, MRR-based topologies provide better energy efficiencies due to their small footprint and thus lower modulation energy consumption. The overall energy efficiency for either of the implementations experience major degradation mainly due to the inefficiency in lasers and phase shifters, the insertion and excess losses of the optical components as well as the optical to electrical and electrical to optical conversion overhead. 

However, an order of magnitude higher operating speed as well as the inherent parallelism in conducting multiplication for analog signals render them attractive for reducing delay and enhancing throughput in comparison to digital CMOS implementations. With the emerging technologies in SiP, e.g. NOEMS and LCOS, low energy tuning schemes have the potential to significantly improve the energy efficiency of the photonic accelerators. Nonetheless, thermal PS with insulation is still an efficient weight-tuning option which is attractive for mass production. Low voltage-swing modulators, with heterogeneous integration of polymers also promise improvements in energy efficiency due to the significant reduction in modulator and CMOS driver power consumption.

Scaling SiP accelerators to larger network sizes is limited by the rated output optical power of the laser as well as the SNR required for a given signal resolution, $n_{i/p}$. MRM-based networks can scale to larger values than their MZM-based counterparts due to the lower loss associated with cascading microrings in a WDM implementation. Incorporating high-power multi-wavelength lasers will be crucial. However, MRR-based networks are more sensitive to temperature, and often require temperature control between the photonic IC and the laser. The size of MRR-based networks that have been demonstrated in prototype hardware have been limited to 8 modulators \cite{TeraPHY2019} or 16$\times$16 switch \cite{Hasitha_Jayatilleka_Photoconductive_heaters_enable_control_of_large_scale_silicon_photonic_ring_resonator_circuits}. Until larger MRM-based networks are demonstrated in hardware, the adoption of MZM-based networks will continue to be favored.

Heterogeneous integration of low-noise SOAs in SiP is a possible way to increase the network size, but the high power consumption of SOAs degrade the energy efficiency significantly. Higher responsivity and low-noise APDs will also prove beneficial in scaling up the network sizes. The size can be further scaled up by reducing the insertion losses of contributors such as directional couplers, phase shifters (if lossy), etc. The splitters remain a significant limitation for scaling. To make efficient use of the limited optical network sizes, general matrix Multiplication (GEMM) algorithms must be adopted. 

Enhancing the energy efficiency can be achieved by adopting modulators with low static and dynamic power consumption, high-responsivity APDs along with TIAs with high sensitivity, and efficient SOAs and lasers with high WPE. To better address the need for high resolution, multi-level signaling significantly beyond PAM4 and PAM8 must be implemented. Controlling the temperature of the chip also helps with maintaining high resolution. Besides, the crosstalk and distortion of SOAs should be further investigated. 


\begin{acknowledgments}
This work was supported by the Natural Sciences and Engineering Research Council of Canada (NSERC). Access to CAD tools and technology is facilitated by CMC Microsystems. The authors acknowledge Dr. Alex Tait at Queen's University and Avilash Mukherjee at UBC for their technical comments. 
\end{acknowledgments}

\section*{Author Declarations}
The authors declare no conflict of interest.

\section*{Data Availability}
The data that supports the findings of this study are available within the article.  

\section*{Appendix: Scaling Up Using SOA\lowercase{s}}\label{appendix_a}
\begin{figure*}
\centering\includegraphics[width=0.810\linewidth]{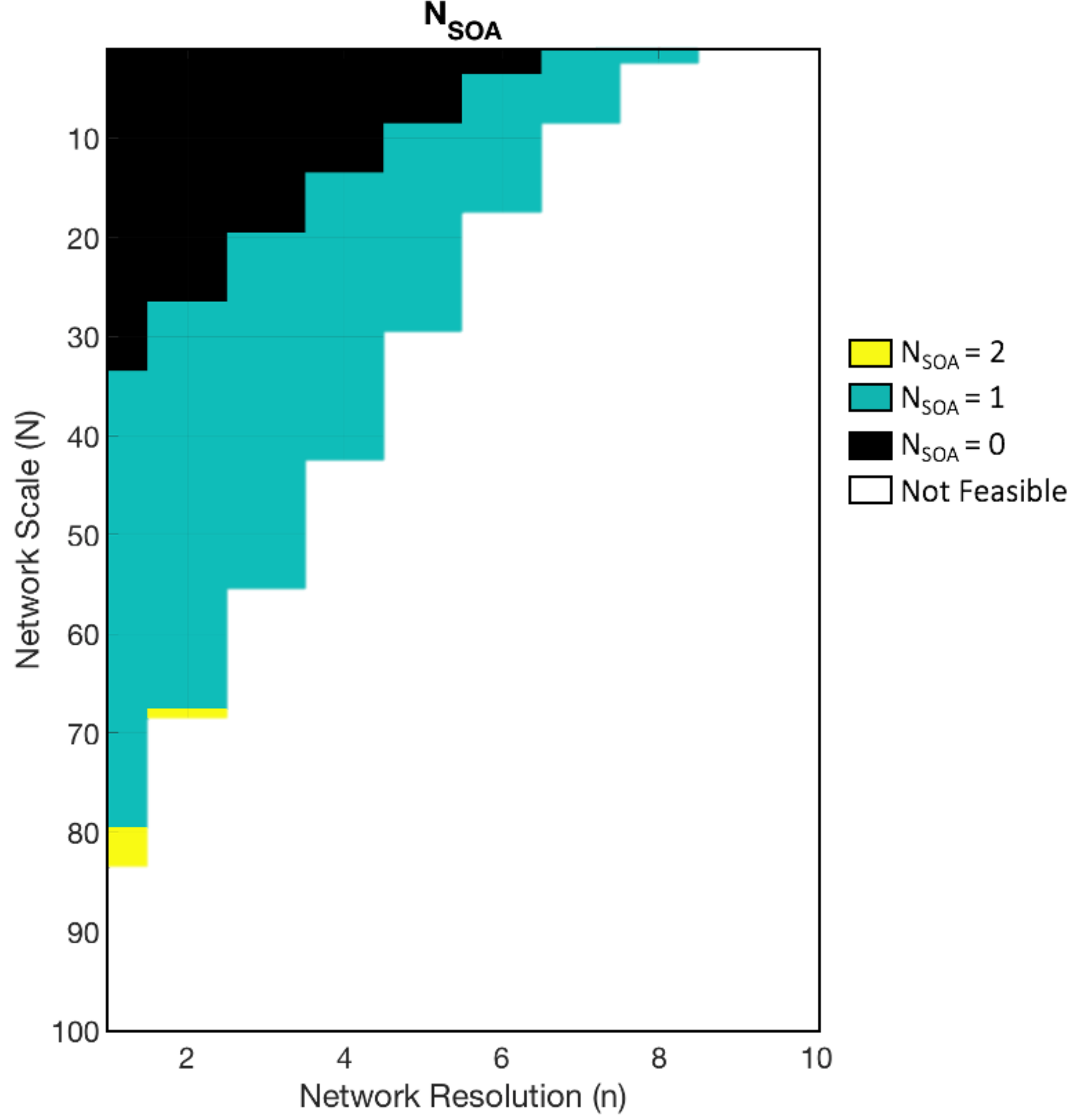}
\caption{\label{fig_MZM_N_vs._n_vs._Nsoa} Network scale vs. resolution for various number of SOAs for an MZM-based implementation. The scale of the network can be traded off with its resolution. Incorporating SOAs help in extending the network scale for a fixed resolution or increasing the resolution for a given network scale. The number of SOAs that can be added to a network are limited to 1 or 2.}
\end{figure*}

\begin{figure*}
\centering\includegraphics[width=0.810\linewidth]{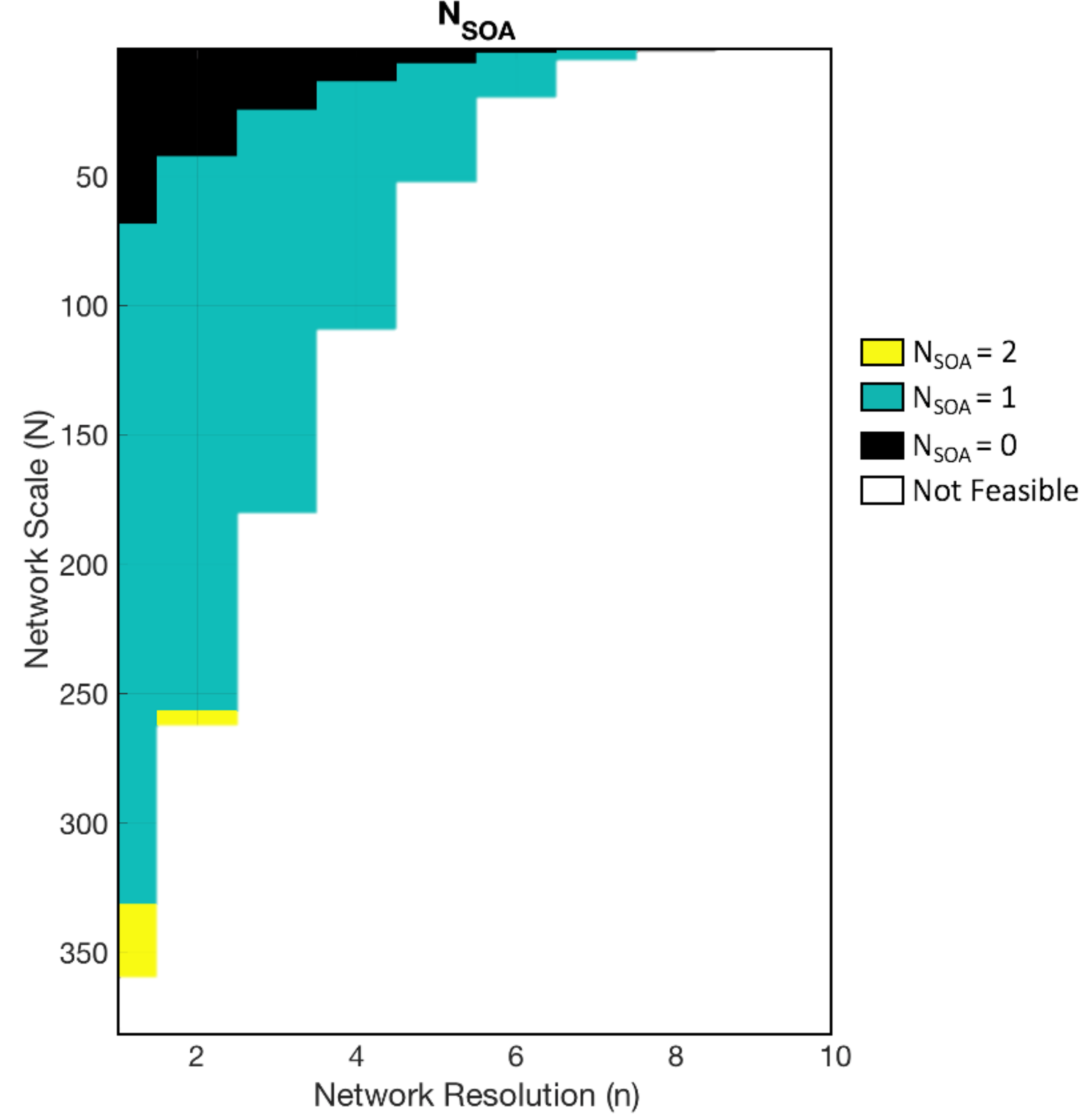}
\caption{Network scale vs. resolution for various number of SOAs for an MRM-based implementation. Compared to their MZM counterpart, SOAs have bigger impact in scaling up the network for a given resolution.}
\label{fig_MRM_N_vs._n_vs._Nsoa}
\end{figure*}

This section illustrates the feasibility of incorporating SOAs in SiP networks in Mach Zehnder and microring based implementations. Fig.~\ref{fig_MZM_N_vs._n_vs._Nsoa} and Fig.~\ref{fig_MRM_N_vs._n_vs._Nsoa} illustrate the number of SOAs in terms of the network resolution and scale for MZM and MRM based implementations, respectively. It can be observed that the resolution which a SiP network is desired to operate at is dependant on the network scale.

It can be inferred that an MZM-based network can support signal resolutions of $4b$ for a network size of $N_{ltd} = 55$, with a single SOA. In comparison, incorporating SOAs in MRM implementations increases the network limited scale to $N_{ltd} = 94$ for $n_{i/p}=4b$ as can be shown in Fig.~\ref{fig_MRM_N_vs._n_vs._Nsoa}. The number of SOAs that can be added to a network are limited to 1 or 2.

\section*{References}
\bibliography{MAC_paper}

\end{document}